\documentclass[traditabstract]{aa}
\usepackage{graphicx,natbib,color}
\usepackage{txfonts,aalongtable,afterpage}

\newcommand{\xspec}{{\sc xspec}}

\newcommand{\nthcomp}{{\sc nthcomp\ }}

\def\lta{ \lower .75ex\hbox{$\sim$} \llap{\raise .27ex \hbox{$<$}} }

\begin{document}

\date{Received .../Accepted ...}

\title{Testing warm Comptonization models for the origin of the soft X-ray excess in AGN}
\author{P.-O. Petrucci\inst{1}
  \and F. Ursini\inst{2}
  \and A. De Rosa\inst{3}
  \and S. Bianchi\inst{4}
  \and M. Cappi\inst{2}
  \and G. Matt\inst{4}
  \and M. Dadina\inst{2}
  \and J. Malzac\inst{5}
  }
  
\institute{Univ. Grenoble Alpes, CNRS, IPAG, F-38000 Grenoble, France 
	   \and
	   INAF-IASF Bologna, Via Gobetti 101, 40129 Bologna, Italy
	   \and 
	   INAF/Istituto di Astrofisica e Planetologie Spaziali. Via Fosso del Cavaliere, I- 00133 Roma, Italy
	   \and
	   Dipartimento di Matematica e Fisica, Universit\`a degli Studi Roma Tre, 
	   via della Vasca Navale 84, 00146 Roma, Italy 
	   \and
IRAP, Universit\'e de toulouse, CNRS, UPS, CNES , Toulouse, France.
} 

%

\abstract{The X-ray spectra of many active galactic nuclei (AGN) show a soft X-ray excess below 1-2 keV on top of the extrapolated high-energy power law. The origin of this component is uncertain. It could be a signature of relativistically blurred, ionized reflection, or the high-energy tail of thermal Comptonization in a warm ($kT\sim$ 1 keV), optically thick ($\tau\simeq$ 10-20) corona producing the optical/UV to soft X-ray emission. 

%
The purpose of the present paper is to test the warm corona model on a statistically significant sample of unabsorbed, radio-quiet AGN with XMM-newton archival data, providing simultaneous optical/UV and X-ray coverage. The sample has 22 objects and 100 observations. We use two thermal comptonization components to fit the broad-band spectra, one for the warm corona emission and one for the high-energy continuum. In the optical-UV, we also include the reddening, the small blue bump and the Galactic extinction. In the X-rays, we include a WA and a neutral reflection.


The model gives a good fit (reduced $\chi^2<$1.5) to more than 90\% of the sample. We find the temperature of the warm corona to be uniformly distributed in the 0.1-1 keV range, while the optical depth is in the range $\sim$10-40. These values are consistent with a warm corona covering a large fraction of a quasi-passive accretion disc, i.e. that mostly reprocesses the warm corona emission. The disk intrinsic emission represents no more than 20\% of the disk total emission. According to this interpretation, most of the accretion power would be released in the upper layers of the accretion flow. 

}


\maketitle

\section{Introduction}
The origin of the soft X-ray excess is a long standing issue in our understanding of the AGN X-ray emission. This excess, above the extrapolation to low energy of the high energy ($>$2 keV) continuum power law fit, has been discovered in the 80's thanks to EXOSAT \citep{arn85} and since then has been observed in a large fraction of AGN (e.g. \citealt{wal93,pag04,gie04,cru06}).

It was realized that its characteristic temperature (when fitted with e.g. a simple black body) 
was remarkably constant over a wide range of AGN luminosities 
and black hole masses (e.g. \citealt{wal93,gie04,bia09a,cru06}), 
favoring an origin through atomic processes instead of purely continuum 
emission. Different models, assuming either blurred ionised reflection  or thermal comptonisation in an optically thick ($\tau>$1) and warm (kT$\sim$1 keV) plasma, have been proposed to fit this component and both give acceptable results \citep{cru06,mag98,don12,jin12c}. \\

The thermal comptonisation modeling of the soft X-ray excess has been carefuly tested with the data set from the large broad band campaign on Mrk 509 (\citealt{kaa11a}, \citealt{pet13}, P13 hereafter). This campaign is still unique in term of duration, energy coverage and number of observations involved. At the core of this program are ten observations with XMM-Newton of approximately 60 ks each spaced by four days. INTEGRAL observations were obtained simultaneously, extending the energy coverage up to the hard X-rays.  A strong correlation between the UV and the soft X-ray ($<$0.5 keV) flux were observed during this campaign, while no correlation was found between the UV and the hard ($>$3 keV) X-rays (see \citealt{mehd11}). This suggested that the UV and soft X-ray emissions were produced by the same spectral component, this interpretation being supported by a Principal Component Analysis (P13). Combined with the absence of a broad iron line component, which would have been expected if blurred reflection would have been present  in this source, the thermal Comptonization appeared naturally as the most plausible scenario. 

The data were then fitted with a ``two-coronae'' model, which assumes two comptonisation model components, one for the UV-Soft X, the so-called warm corona, and one for the hard X, the so-called hot corona (see P13 for more details). Warm absorption and neutral reflection were also included in the fit procedure. The model gave statistically good fits for all the observations of the campaign and it provided very interesting information on each possible corona geometry. The disk-hot corona system agrees with a ``photon-starved'' configuration, i.e. a disk-corona geometry where the solid angle under which the corona ``sees'' the accretion disk is small. This is a common result for Seyfert galaxies since it is known that large covering factor coronae (e.g. slabs) cannot reproduce (due to the strong compton cooling on the disk soft photons) hard X-ray spectra like those observed in Seyfert galaxies (e.g. \citealt{haa93,haa94}). In contrast, the analysis of P13  shows that the {warm} corona agrees very well with a powerful, extended and optically thick plasma covering a passive accretion disk, i.e., all the accretion power would be released in the warm corona! This is a result at odd with the commonly accepted behavior of standard {  optically thick, geometrically thin} accretion flows \citep{sha73}, where the gravitational power is believed to be released in the deeper layers. \\

If true, the consequences are important, with direct impact on e.g. our understanding of the accretion disk vertical equilibrium, the expected spectral emission from such accretion flow, its capacity of producing outflows/jets, etc... \cite{roz15} did a theoretical study of the existence of such a warm optically thick corona  at the surface of a standard accretion disk. By varying the accretion power released into the corona with respect to the underlying disk, as well as the magnetic pressure, this study shows that it is indeed possible to obtain solutions having the required temperature and optical depth (see their Fig. 4). Specifically, the best conditions are obtained when the majority of the power of accretion is released into the corona rather than the disk, in good agreement with the conclusion obtained with the Mkn 509 campaign. Large magnetic to gas pressure is needed ($>$30) to reach such large optical depth (10-20) while still ensuring hydrostatic balance \footnote{The presence of outflows would relax this constraint however.}.\\

The purpose of the present paper is to test more consistently the warm corona model interpretation for the soft X-ray excess in radio quiet AGNs.  We present in Sect. \ref{thermcompt} the general equations governing the radiating equilibrium in a disk-corona structure. We show how these equations can be used to extract the physical properties of any corona (assuming its emission is dominated by thermal comptonisation) from the spectral fit parameters of its high energy emission (mainly its photon index $\Gamma$ and  cut-off temperature $kT$). Then,  in Sect. \ref{softXlit}, we compare these theoretical expectations with the different spectral analyses of the soft X-ray excess done in the literature and show their agreement with a quite extended corona geometry above a passive disk. Yet, none of these published spectral analysis consistently test the ``two-coronae'' model  to a large sample of AGN. This is the goal of Sect. \ref{test2corona} where we describe our sample selection, our methodology and our results before concluding. 


\section{Disk-Corona radiative equilibrium}
\label{thermcompt}
\subsection{Main equations}
\label{maineq}
\begin{figure*}[t]
\begin{center}
\includegraphics[width=0.8\textwidth]{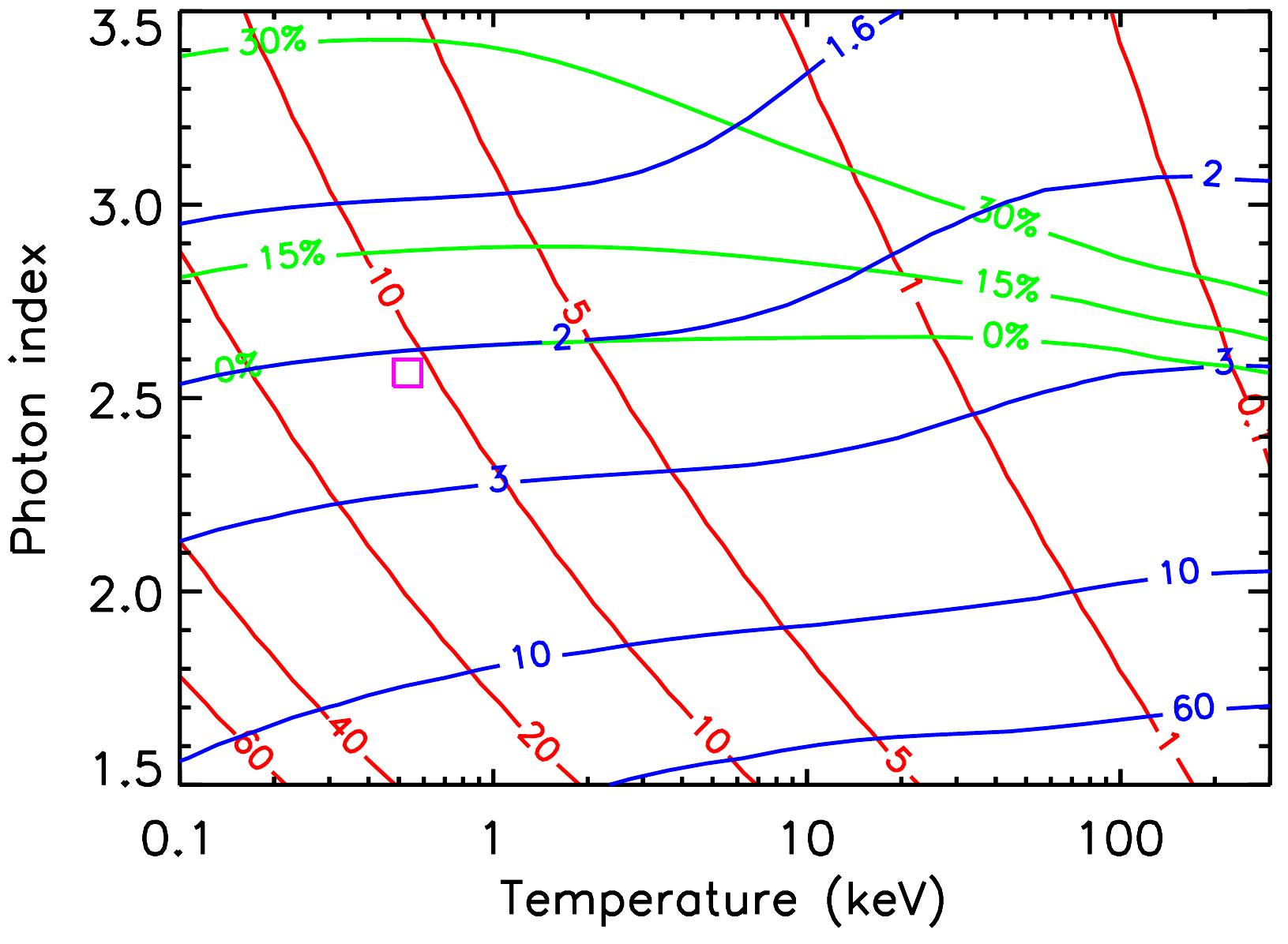}
 \caption{Contours of the corona optical depth $\tau$ in red, the amplification factor $A$ (in blue) and the minimal fraction (in \%) of disk intrinsic  emission $\displaystyle \left . \frac{L_{disk,intr}}{L_s}\right |_{min}$ (in green)  in the corona spectral parameter $\Gamma$-$kT$ plane. A slab geometry (i.e. $g=1$ in Eq. \ref{eqdiskint}) is assumed here. Smaller values of $g$ (i.e. a patchy corona) would move the green contours down. The magenta rectangle corresponds to the $\Gamma$ and temperature $kT_e$ ranges obtained by P13 for the soft excess in the spectral analysis of the Mkn 509 campaign.
 }
\label{figgamkt}
\end{center}
\end{figure*}

In a disk-corona system, the Comptonizing region (i.e. the corona) and the source of soft photons (i.e. the disk) are coupled, as the optically thick disk necessarily reprocesses and reemits part of the Comptonized flux as soft photons which are the seeds for Comptonization. The system must then satisfy equilibrium energy balance equations, which depend mainly on the disk-corona geometry 
(see e.g. \citealt{haa91,ste95}). We develop a little in this section what these energy balance equations imply in terms of observables (cf. Appendix of \cite{pet13} for more details. {  See also Appendix B of \cite{kub16}}).\\ 

{  We assume the accretion disk be mainly neutral, its temperature being rather low in AGNs.} From the observed comptonized spectrum, we can deduce an observed photon rate $\dot{n}_{obs}$. By conservation of the number of photons, which characterizes the Compton process\footnote{We neglect the pair creation/annhilation process here which is a reasonable assumption for low ($<$500 keV) temperature plasma. {  We also neglect the photon rate from the reflection component. For neutral accretion disk, it is negligible in comparison to the disk photon rate.}}, this photon rate is equal to the sum of the seed photons rate crossing the corona without being comptonized $\dot{n}_{s,0}$ and those comptonized in the corona and emitted upward (in direction to the observer) $\dot{n}_{c,up}$. At first order,
\begin{eqnarray}
\dot{n}_{s,0}&=&\dot{n}_se^{-\tau}\label{eqns1}\\
\dot{n}_{c,up}&=&\frac{\dot{n}_s(1-e^{-\tau})}{2},\label{eqns2}
\end{eqnarray}
where $\dot{n}_s$ is the photon rate emitted by the disk that enters the corona and the factor  $1/2$ of the last equation assumes the Compton scattering process to be isotropic  with half of the comptonized photons being emitted upward, and the other half being emitted downward. The Compton process is certainly not isotropic, especially in the case of an anisotropic seed soft photon field (e.g. \citealt{haa91,haa93,ste95,hen97}), but the effect on the photon rate is relatively small.\\

In consequence, we have 
\begin{equation}
\dot{n}_{obs}=\dot{n}_{s,0}+\dot{n}_{c,up}=\frac{\dot{n}_s(1+e^{-\tau})}{2},
\end{equation}
which gives
\begin{equation}
\dot{n}_{s}=2\frac{\dot{n}_{obs}}{(1+e^{-\tau})}\label{nsnobs}.
\label{dotn}
\end{equation}
For an optically thin corona ($\tau\ll 1$), $\dot{n}_s\simeq \dot{n}_{obs}$ (very few soft photons are comptonized). On the other hand, for $\tau\gg 1$, $\dot{n}_s\simeq 2 \dot{n}_{obs}$. Then, once the photon rate $\dot{n}_s$ of the accretion disk crossing the corona is known, as well as the accretion disk temperature (deduced from the fit), we have access to the luminosity $L_s$ coming from the disk that cools the corona.\\

On the other hand, the corona total power $L_{tot}$ is the sum of $L_s$, the heating power $L_h$ liberated in the corona:
\begin{equation}
L_{tot}=L_h+L_s.\label{eqtot}
\end{equation}
Here, $L_h$ and $L_s$ can be divided into upward and downward:
\begin{eqnarray}
L_h&=&L_{h,u}+L_{h,d}=2L_{h,u}\label{eqlhup}\\
L_s&=&L_{s,u}+L_{s,d}\\
&=&\underbrace{L_se^{-\tau}+\frac{L_s(1-e^{-\tau})}{2}}_{L_{s,u}} + \underbrace{\frac{L_s(1-e^{-\tau})}{2}}_{L_{s,d}}\label{eqlsup}
\end{eqnarray}
The first equation assumes an isotropic Compton process, and the second equation is obtained with the same reasoning as in Eq. \ref{eqns1} and \ref{eqns2}. \\

Then, {  being $L_{ref}$ the reflected luminosity at the disk surface}, the observed luminosity can be written:
\begin{equation}
L_{obs}=L_{h,u}+L_{s,u}  {+L_{ref}}.\label{eqobs}
\end{equation}
{  $L_{ref}$ is directly related to the down-scattered corona emission}:
\begin{eqnarray}
  {L_{ref}} &=&   {a(L_{s,d}+L_{h,d})}\\
&=&  {a\left (\frac{L_s(1-e^{-\tau})}{2}+\frac{L_h}{2}\right )}
\label{eqlrefup}
\end{eqnarray}
{  where $a$ is the disk albedo. Since we assume the disk to be neutral, the albedo is not expected to be large. As shown by Haardt \& Maraschi (\citeyear{haa93b}, see their Fig. 2), in a radiatively-coupled disc-corona system the albedo depends on the corona optical depth and is of the order of 10\% for small corona optical depth $\tau\sim$0.1. It rapidly decreases to 0 as $\tau$ increases since the corona spectrum softens and is efficiently absorbed by the neutral disk matter. We take into account the albedo dependency with $\tau$ by digitalizing and interpolating the function $a(\tau)$ plotted in Fig. 2 of \cite{haa93b} and assuming $a = 0$ for $\tau >1$.}\\

 From Eqs. \ref{eqlhup}, \ref{eqlsup}, and \ref{eqobs}, we can deduce $L_h$
\begin{equation}
L_h=2L_{obs}-{L_s}(1+e^{-\tau})  {-2L_{ref}}\label{eqLh},
\end{equation}
and finally, from Eq. \ref{eqtot}
\begin{equation}
L_{tot}=2L_{obs}-{L_s}e^{-\tau}  {-2L_{ref}}\label{eqLtot}.
\end{equation}
Let us introduce the comptonisation amplification factor $A$ defined by $L_{tot}=AL_s$. 
From the above equations we deduce that:
\begin{eqnarray}
A&\simeq& 2L_{obs}/L_s  {-2L_{ref}/L_s}-1,\mbox{ for $\tau\ll 1$}\label{eqAa}\\ 
A&\simeq& 2L_{obs}/L_s  {-2L_{ref}/L_s}, \mbox{ for $\tau\gg 1$}\label{eqAb}
\end{eqnarray}

The radiative equilibrium of the disk also implies
\begin{equation}
  { L_{disk} = (1-a)(L_{s,d}+L_{h,d})+L_{disk,intr}.}
\end{equation}
{  where $  {(1-a)(L_{s,d}+L_{h,d})}$} is the corona emission reprocessed in the disk and $L_{disk,intr}$ is the intrinsic (i.e. locally generated in the disk) disk emission. Concerning the luminosity $L_s$, it characterizes the part of the disk emission $L_{disk}$ entering and cooling the corona. It is equal to $L_{disk}$ if the corona covers entirely the accretion disk. But it is lower than $L_{disk}$ in the case of e.g. a patchy corona where part of the disk emission reaches the observer without crossing the corona. So we have
\begin{eqnarray}
  {L_s} &<&   {(1-a)(L_{s,d}+L_{h,d})+L_{disk,intr}}\\
&=&   {g\left [ (1-a)\left (\frac{L_s(1-e^{-\tau})}{2}+\frac{L_h}{2}\right )+L_{disk,intr}\right ]}
\end{eqnarray}
where $g$ is a geometrical parameter, $\leq$1, related to the patchiness of the corona. Geometrically, it  gives an estimate of the solid angle $\Omega$ under which the disk ``sees'' the corona, i.e. $g=\Omega/2\pi$. In consequence, the heating/cooling ratio is equal to
\begin{equation}
  {\frac{L_h}{L_s}=A-1=\frac{2}{1-a}\left (\frac{1}{g}-\frac{L_{disk,intr}}{L_s}\right )+(e^{-\tau}-1)}.
\label{eqratio}
\end{equation} 
which gives, thanks to Eqs. \ref{eqlrefup} and \ref{eqLh},
\begin{eqnarray}
\frac{L_{disk,intr}}{L_s} &=&   {\frac{1}{g}+\frac{2(1-a)}{2+a}\left (e^{-\tau}-\frac{L_{obs}}{L_s}\right )}\\
&\ge&   {1+\frac{2(1-a)}{2+a}\left (e^{-\tau}-\frac{L_{obs}}{L_s}\right )}=\left . \frac{L_{disk,intr}}{L_s}\right |_{min}.
\label{eqdiskint}
\end{eqnarray} 
In the case of an optically thin ($\tau \ll1$) corona entirely covering ($g$=1) a passive ($L_{disk,intr}$=0) and non reflective ($a$=0) disk, we found the well known result $L_{h}/L_s=2$ (see e.g. \citealt{haa91} and their Eq. 3b with f=0 and no disk albedo) and consequently the amplification ratio $A=L_{tot}/L_s=3$. On the other hand, for an optically thick corona (still entirely covering  a passive disk with no albedo), we find $L_{h}/L_s=1$ and $A=L_{tot}/L_s=2$. Larger values of $A$ compared to these fiducial values necessarily required $g<1$ i.e. a patchy corona. On the other hand, lower values of $A$ imply $L_{disk,intr}>$0.\\



\subsection{Mapping}
For a given geometry, equations \ref{eqLh}, \ref{eqLtot} and \ref{eqdiskint} link together the model parameters ($L_{obs}$ and $\tau$) to the intrinsic characteristics (A=$L_{tot}/L_s$ and $\displaystyle\left . \frac{L_{disk,intr}}{L_s}\right |_{min}$) of the corona-disk system. It is then possible to ``map" one pair of parameters in the 2D-plane of the other pair. This can be done as follows. For this purpose we used the comptonisation model {\sc nthcomp} \citep{zdz96,zyc99} in \xspec \citep{arn96,arn99}. {\sc nthcomp} is characterized by three main parameters, the corona electron temperature $kT_e$, the disk photon temperature $kT_{bb}$ (we assume a multicolor disk blackbody) and the power law photon index $\Gamma$ of the Comptonized spectrum. Then we proceed through the following steps. 
\begin{itemize}
\item We first assume a disk photon temperature $kT_{bb}$. Let us assume 3 eV but the result is however weakly dependent on this parameter as long as it is of order of a few eVs.
\item we choose a set of model parameter values $\Gamma$, $kT_e$ and produce a SED.
\item from $\Gamma$ and  $kT_e$ we estimate the corresponding optical depth $\tau$ using eq. Eq. 13 of \cite{bel99}\footnote{This is an approximative estimate of $\tau$, but it is sufficient for our current purposes}, i.e. $\displaystyle  {\Gamma\simeq\frac{9}{4}y^{-2/9}}$ with $  {y=4 [kT_e/m_ec^2+4(kT_e/m_ec^2)^2)]\tau(\tau+1)}$ {  the so-called Compton parameter}. From the SED integration, we also compute $L_{obs}$ and $\dot{n}_{obs}$ \footnote{The unit of $L_{obs}$ and $\dot{n}_{obs}$ do not matter here because we deal with luminosity and photon rates ratios}.
\item then we estimate $L_s$ for the given $T_{bb}$ by assuming photon conservation in slab geometry (Eq. \ref{nsnobs})
\end{itemize}
Following this procedure, and for a given $kT_{bb}$, we can compute $\tau$, $L_{obs}$ and $L_{s}$, or similarly, given Eq. \ref{eqAa}, \ref{eqAb} and \ref{eqdiskint}, $\tau$, $A$ and  $\displaystyle\left .\frac{L_{disk,intr}}{L_s}\right |_{min}$ in function of $\Gamma$ and $kT_e$. We have reported in Fig. \ref{figgamkt} 
the corresponding contours of the optical depth $\tau$ (in red), the amplification factor $A$ (in blue) and the minimal fraction of disk intrinsic emission $\displaystyle\left . \frac{L_{disk,intr}}{L_s}\right |_{min}$ (in green)  in the ($\Gamma$, $kT_e$) plane for $\Gamma$ varying between 1.5 and 3.5 and $kT_e$ between 0.1 keV and 300 keV. \\

The contour $\displaystyle\left . \frac{L_{disk,intr}}{L_s}\right |_{min}=0\%$ corresponds to the theoretical case of a slab corona above a passive disk. This contour divides Fig.  \ref{figgamkt} in two parts. Above it, a slab corona-disk geometry can exist, the farther up the larger the disk intrinsic emission. On the other hand, below it the disk-corona geometry is necessarily patchy. Note also that at large optical depth (left part of the Figure), this contour agrees with an amplification factor of 2 while at low optical depth (right part of the Figure) it agrees with $A$=3, as theoretically expected.

Since observations provide us with $\Gamma$ and $kT_e$, we are able to deduce, for a given object, the corresponding  values of $\tau$, $A$ and  $\displaystyle\left . \frac{L_{disk,intr}}{L_s}\right |_{min}$.\\

{  Interestingly, the contours $\displaystyle\left . \frac{L_{disk,intr}}{L_s}\right |_{min}=0\%$ and $A=2$ correspond, for large optical depths, to corona  photon index $\Gamma\sim$ 2.5-2.6. This is indeed the photon index values for which $\dot{n}_s\simeq 2\dot{n}_{obs}$  and $L_s\simeq L_{obs}$ (Eqs. \ref{dotn} and \ref{eqAb} respectively with $\tau\gg$1) are satisfied (see details in App. \ref{app1}). This corresponds also to the situation $L_h=L_s$ i.e. a corona heating of the order of the disk luminosity.}\\

\section{Indication of a dominantly dissipating and extended warm corona for the origin of the soft X-ray excess}
\label{softXlit}
As said in the introduction, the ``two-coronae'' model was carefully tested with the optical/UV/X-ray data set from the broad band campaign on Mrk 509 (P13). This spectral analysis showed that the warm corona spectral emission was quite well constrained with a photon index $\Gamma$ and temperature $kT_e$ varying in the ranges 2.53--2.6 and 0.48--0.59 keV respectively. Looking at Fig. \ref{figgamkt}, and conformably to the conclusions of P13, these parameter values agree with an optically thick corona ($\tau$ between 10 and 20), in a slab geometry ($A\simeq$2) above a passive disk $\left(\displaystyle\left . \frac{L_{disk,intr}}{L_s}\right |_{min}\simeq 0\right )$.\\

Interestingly, past spectral analyses of the soft X-ray excess on large sample of objects show similar values for $\Gamma$ and, when included in the modeling, of $kT_e$. From their analysis of the 0.1--2.4 keV ROSAT spectra of 58 Seyfert 1, \cite{wal93} found photon indices varying between 2.0 and 3.2 with a mean of 2.5 (see their Fig. 2). Similar results for $\Gamma$ were obtained by \cite{bru97}, \cite{lao97} or \cite{sch96}  from their analysis of the ROSAT spectra of different quasar samples. More recently, \cite{broc06} analysed  XMM-Newton EPIC-pn and OM observations for 22 Palomar Green quasars. They obtained a reasonable fit of the 0.3--10.0 keV spectra of most of the sources with a broken power law (plus an iron line), the low-energy power law (to fit the soft excess) photon index falling in the 2.0--4.0 range, with an average value around 2.9.

A few years ago, \cite{jin12c} presented the broad band (from optical to  hard X-rays) spectral analysis of a sample of 51 unobscured Type 1 AGN, fitting together XMM-Newton and Sloan Digital Sky Survey spectra with a model very similar to the ``two-coronae" one. It assumes that the gravitational potential energy is emitted as optically thick blackbody emission at each radius down to some specific coronal radius. Below this radius the remaining energy down to the last stable orbit is divided between two coronae that play the same roles of the warm and hot coronae of the ``two-coronae" model {  (the slab geometry is also discussed in the appendix of \citealt{don12}). The main difference with our approach, is that their model assumes, by construction, intrinsic disk emission.} 
Their spectral analysis show that the warm corona temperatures and optical depths\footnote{Their comptonisation model for the warm corona is {\sc comptt} in \xspec. {\sc comptt} has the temperature and optical depth  as corona parameter} of the whole sample cluster around 0.2-0.3 keV and 10-20 respectively. While these parameter values theoretically agree with their assumptions of a patchy warm corona above an intrinsically radiative disk (indeed for $g<1$ the green contours of $\displaystyle\left .\frac{L_{disk,intr}}{L_s}\right |_{min}$ move down in Fig. \ref{figgamkt}), it is quite surprising that they also agree very well with a slab corona above a passive disk.\\ 

These studies show that, whatever the model used for the soft X-ray excess, the data constrain the spectral parameters to fall in the region of the $\Gamma-kT$ space that is consistent with a dominantly dissipating and extended warm corona, in good agreement with the conclusions of P13 for Mkn 509. Now, most of the past works used only a phenomenological power law model to fit the soft X-ray excess but did not include the optical-UV data. On the other hand, the detailed spectral analysis of \cite{jin12c} do include the optical-UV and Soft X-ray data and use realistic comptonisation modeling, but their model assume intrinsically that a significant part of the optical-UV is directly produced by the accretion disk.\\ 

The goal of the following section is to test the results obtained on Mkn 509 on a larger sample of radio quiet AGNs by fitting optical-UV, soft X-ray up to hard X-rays data (10 keV) with the ``two-coronae'' model. Contrary to past analyses, in the ``two-coronae model'', the optical/UV to Soft X-ray emission is entirely due to the warm corona. We want to check, first, if this model can  fit reasonably well other AGNs and, second, see how the warm corona best fit parameters compare with the theoretical expectations discussed in Sect. \ref{thermcompt}.

\section{Testing the ``two-coronae'' model on a sample of AGN}
\label{test2corona}
\subsection{Sample selection and data reduction}
\label{obs}
To test the ``two-coronae" model we need bright sources with simultaneous data in the optical/UV and X-ray bands. Simultaneity is important here because these sources are generally variable at all wavelength on day/weeks timescales (e.g. \citealt{pon12a}).  The capabilities of XMM-Newton are optimal for this task, as it provides both high-quality X-ray spectra with the EPIC-pn camera \citep{stru01}, 
and optical/UV data with the optical monitor (OM hereafter, \citealt{mas01}). Our sample was thus built starting from the sources observed by XMM-Newton, with public data as of April, 2014, cross-correlated with the AGNs and quasars catalogue of \cite{ver10}.
The sources were further selected using the criteria of the CAIXA catalogue \citep{bia09a}, i.e. the sources are radio-quiet and unobscured ($N_H < 2\times 10^{22}$ cm$^{-2}$)\footnote{At this selection step, sources like 1H 0707-495 and Mrk 766 have been discarded from the sample due to their known complex absorption/reflection components (e.g. \citealt{Lgal04,sak03}).}. 

In addition to this, and in order to have better constraints in the optical/UV range during the fitting procedure, we want  the largest number of OM-filters for each source of the sample. Indeed, the variability (in flux and shape) of the sources in the OM band is expected to be small (e.g. \citealt{gel15}) so that the different OM-filter measurements could be used simultaneously to give reasonable constraints to the disk emission (see Sect. \ref{fitproc} for more details). For this purpose, we first cross-correlate the initial sample with the XMM-Newton/OM serendipitous UV source survey catalogue (OMSUSS v2.1, \citealt{pag12}). We then select the largest number of observations per source in order to have at least four OM-filter detections.  
The complete sample of sources and corresponding ObsIDs is listed in Table \ref{tab1}. It corresponds to 22 objects and 100 ObsIDs.\\

The data reduction is identical to the one detailed in \cite{bia09a} i.e. the EPIC-pn data were all reprocessed with the most updated versions of the SAS software \citep{gab04}. 
For the observations performed in Small Window mode, background spectra were generated using blank-field event lists, according to the procedure presented in \cite{rea03}. 
 In all other cases, background spectra were extracted from source-free regions close to the target in the observation event file. Source extraction radii and screening for intervals of flaring particle background were performed via an iterative process that leads to a maximization of the signal-to-noise ratio, similar to what described in \cite{pic04}. 
 Spectra were binned in order to oversample the intrinsic instrumental energy resolution by a factor not lower than 3 and to have spectral bins with at least 25 background-subtracted counts. This ensures the applicability of the $\chi^2$ statistics.


\subsection{Methodology}
\label{meth}
\subsubsection{The model components}
\label{model}
We describe in this Section the different model components we use to fit our sample of optical/UV-X-ray spectra. Since we are mostly interested in the continuum spectral parameters, we do not need a very precise description of the data and we make the choice of fitting with a limited number, but physically motivated, spectral components. The main assumption of the ``two-coronae" model is that the optical/UV emission is the signature of an optically thick multi-color accretion disk whose optical/UV photons are comptonised in two different media: the first one (the so-called warm corona) producing the soft X-ray excess and the second one (the so-called hot corona) the high-energy continuum emission. The different model components described below are also reported in Tab. \ref{tabparam}. We also indicate in this table the parameters of each component that are let free to vary (or not) during the fitting procedure.

\vspace*{-0.25cm}
\paragraph{Continuum:}
We used the \nthcomp model of \xspec to model each of the corona comptonisation emission. As said before, the free parameters of \nthcomp are the electron temperature of the corona $kT_e$, the soft-photon temperature $kT_{bb}$ and the asymptotic power-law index $\Gamma$. In the following 
we will use the subscripts ``wc'' for the warm corona parameters and ``hc" for the hot corona ones. Due to the lack of data above 10 keV, the temperature of the hot corona is fixed to 100 keV. We also assume the same soft-photon temperature for the two coronae.\\

\vspace*{-0.25cm}
\paragraph{Reflection component:}
We include a reflection component (the  {\sc xillver} model in \xspec, \citealt{gar10}) 
leaving only the normalization as a free parameter. Indeed, the main feature of {\sc xillver} in the 0.3-10 keV band is the Fe K$_{\alpha}$ line, plus some contribution to the continuum above 8 keV. Therefore, the spectral shape of the illuminating radiation of {\sc xillver} is not crucial. We thus fixed the photon index of {\sc xillver} to the arbitrary value of 1.9 and assumed an iron abundance of 1 and an ionization parameter log $\xi$ = 0.  We checked that linking the photon index of {\sc xillver} to that of the primary continuum does not alter much the results. We also fixed the power law cut-off energy to $E_c=3kT_e=$300 keV \citep{pet01}.

\vspace*{-0.25cm}
\paragraph{Neutral and warm absorption:}
Concerning the presence of absorption, we first assume a Galactic neutral hydrogen column density $N_h$ ({\sc tbabs} model of \xspec), appropriate for the sky coordinates of the source, from \cite{kalb05}. 
{  $N_h$ is fixed during the fitting procedure.}
We add a WA component modeled with a CLOUDY table, leaving free the column density and the ionization parameter (``cloudytable" model component in Tab. \ref{tabparam}. See \cite{cap16} for more details on this CLOUDY table.).

\vspace*{-0.25cm}
\paragraph{Reddening, small Blue Bump and Galactic emission:}
Since we use optical/UV data, we must take into account two main components that can give a significant contribution in the optical/UV band. One is the host galaxy (``galaxy" model component in Tab. \ref{tabparam}), especially in the optical band, its UV emission being expected to be negligible. The second main component is the Broad Line Region, which is responsible for the so-called small blue bump around 3000 \AA\ (``smallBB" model component in Tab. \ref{tabparam}). We refer the reader to \cite{mehd15} 
 for a detailed description of these contributions in the case of the NGC 5548 XMM-NuSTAR campaign \citep{kaa14}. For the galaxy contribution, we assumed the same template spectrum as the one used by \cite{mehd15} for NGC 5548, the precise spectral shape being of weak importance when using only six broad-band filters. Finally, we included the reddening ({\sc redden} model in \xspec), calculated from the Galactic extinction following \cite{guv09}. {  The redenning is fixed during the fitting procedure.} 

\noindent The complete model we used, in {\sc xspec}  terms, is then:
\begin{eqnarray}
& & tbabs \times redden \times mtable\{cloudytable\} \times ( nthcomp _{wc}+ \nonumber \\
& &  nthcomp_{hc} + atable\{smallBB\} +atable\{galaxy\} + \nonumber \\
& & xillver) \nonumber\\
\end{eqnarray}
This model has 9 free parameters (see Tab. \ref{tabparam}). For each source, the redshift is taken from the NASA/IPAC Extragalactic Database (NED) or from the SIMBAD database operated at CDS, Strasbourg.\\

In several cases however, our best fits show residuals of emission/absorption line-like features below 1 keV and around 2 keV. Albeit rather weak (typically a few eV equivalent width), they can be statistically significant depending on the statistics of the data. Part of the residuals are due to calibration issues (see \citealt{cap16} for a detailed discussion). Those around 2 keV are probably ascribed to remaining systematic calibration uncertainties owing to the detector quantum efficiency at the Si K-edge (1.84 keV) and mirror effective area at the Au M-edge ($\sim$2.3 keV). It was then decided to cut the 1.8-2.4 keV part of the spectrum out. Features at energies lower than $\sim$1.5 keV could be modeled by a combination of a few narrow absorption and/or emission lines at energies around $\sim$0.5-0.6 keV and 1-1.1 keV, and EW variable between $\sim$8-15 eV, depending on the line and observation considered. We estimate that the origin of these features could be ascribed to either remaining uncertainties in the CTI-energy scale at low energies in the pn data, or to an improper (or approximate) modeling of the emission and absorption lines.

Another part of the residuals can also be due to a bad modeling of the WA features. The goal here however is not to obtain a precise fit of the WA. We are rather interested by the continuum and we believe that our results are not significantly impacted by the WA modeling. We discuss this point in more detail in Sect. \ref{goodfit}. This is also supported by the fact that our methodology applied to Mkn 509 give results in good agreement with those obtained by P13 where the WA was precisely taken into account.   In some cases however, and if the improvement is indeed significant, we added another WA component (see next section).\\

\begin{table}
\begin{tabular}{ccc}
Model component$^1$ & Free parameters & Fixed parameters \\
in xspec &  &  \\
\hline
\hline
tbabs & - & $N_h$ \vspace*{0.2cm}\\
redden & - & E(B-V)\vspace*{0.2cm}\\
cloudytable & $N_h, \xi$ & $v_{turb}=100$\vspace*{0.2cm}\\
smallBB & - & norm\vspace*{0.2cm}\\
galaxy & - & norm\vspace*{0.2cm}\\
nthcomp$_{wc}$ & $\Gamma_{wc}$, T$_{e,wc}$ & - \\
 & $T_{bb,wc}, norm_{wc}$ & \vspace*{0.2cm}\\
nthcomp$_{hc}$ & $\Gamma_{hc},  norm_{hc}$ & $T_{e,hc}=100 keV$\\
 &  & $T_{bb,hc}=T_{bb, wc}$\vspace*{0.2cm}\\
xillver & norm & $\Gamma=1.9, \xi=1$\\
 &  & $E_c=300$ keV\\
 &  & $i=30^{\circ}, A_{Fe}=1$ \\
\hline
\end{tabular}

\caption{\label{tabparam} Model component and parameters of the ``two-coronae'' model described in Sect. \ref{model}. The normalisations of the ``smallBB" and the ``galaxy" templates are fixed to their best fit values obtained when fitting the OM data only (see Sect. \ref{fitproc} for more details on the fitting procedure) }

\end{table}

It is important to point out that no link between the Comptonized spectrum and the soft UV emission is imposed a priori in our modeling. The ``two-coronae" model simply adjusts its parameters, independently of each other, to fit the data. It is only a posteriori that the resulting best-fit values of the coronae characteristics can be interpreted in a physically motivated scenario.

\begin{figure}
\includegraphics[width=0.9\columnwidth]{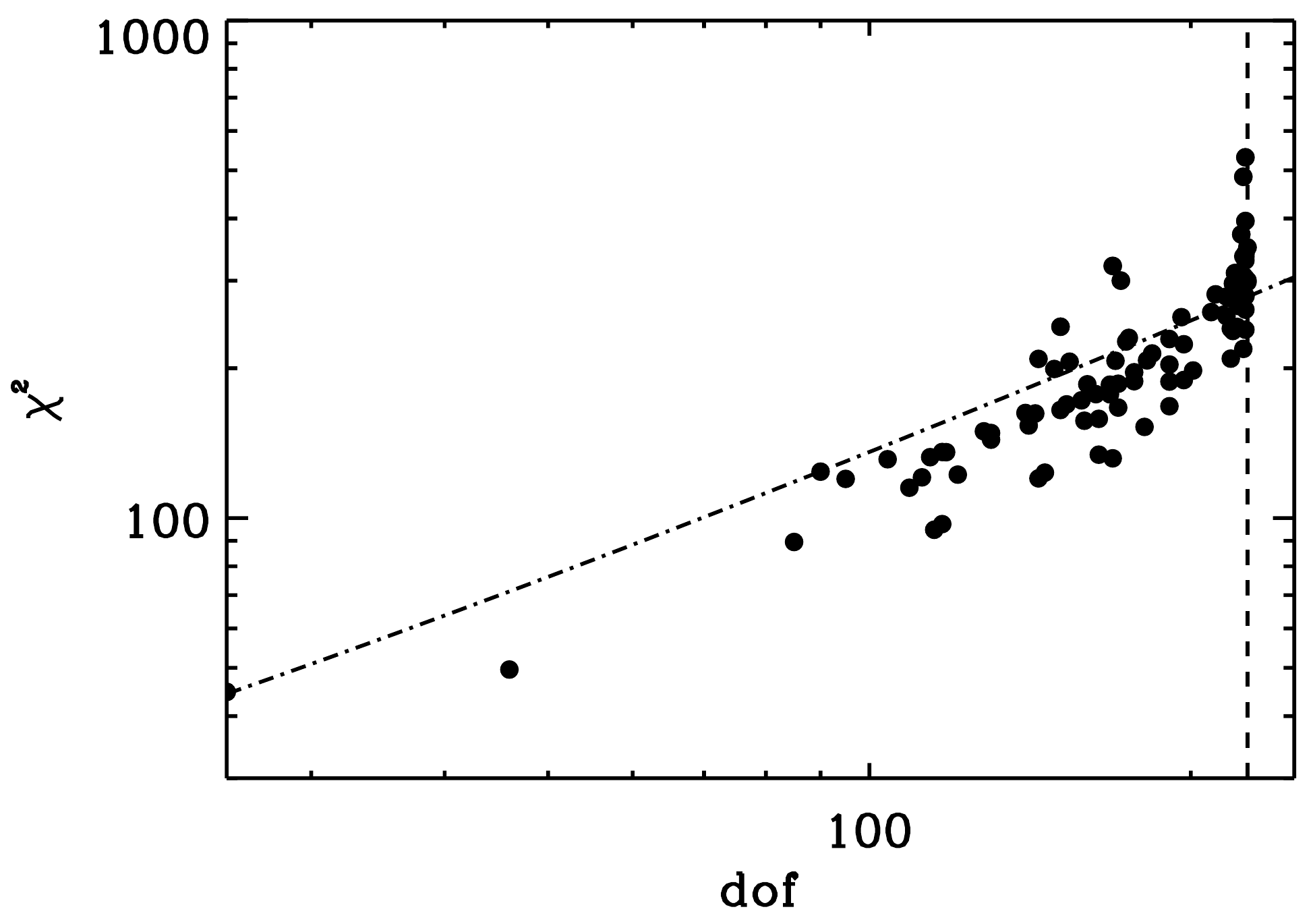}
 \caption{Distribution of our best fit $\chi^2$ in function of the degree of freedom (dog) of each observation of our sample. 
The dot-dashed curve indicates the $\chi^2$ value corresponding to a 99\% null hypothesis probability for the corresponding dof value.The vertical dashed line represents the 226 dof limit of our sample.}
\label{chi2dist}
\end{figure}
\begin{figure}
\begin{center}
\includegraphics[width=\columnwidth]{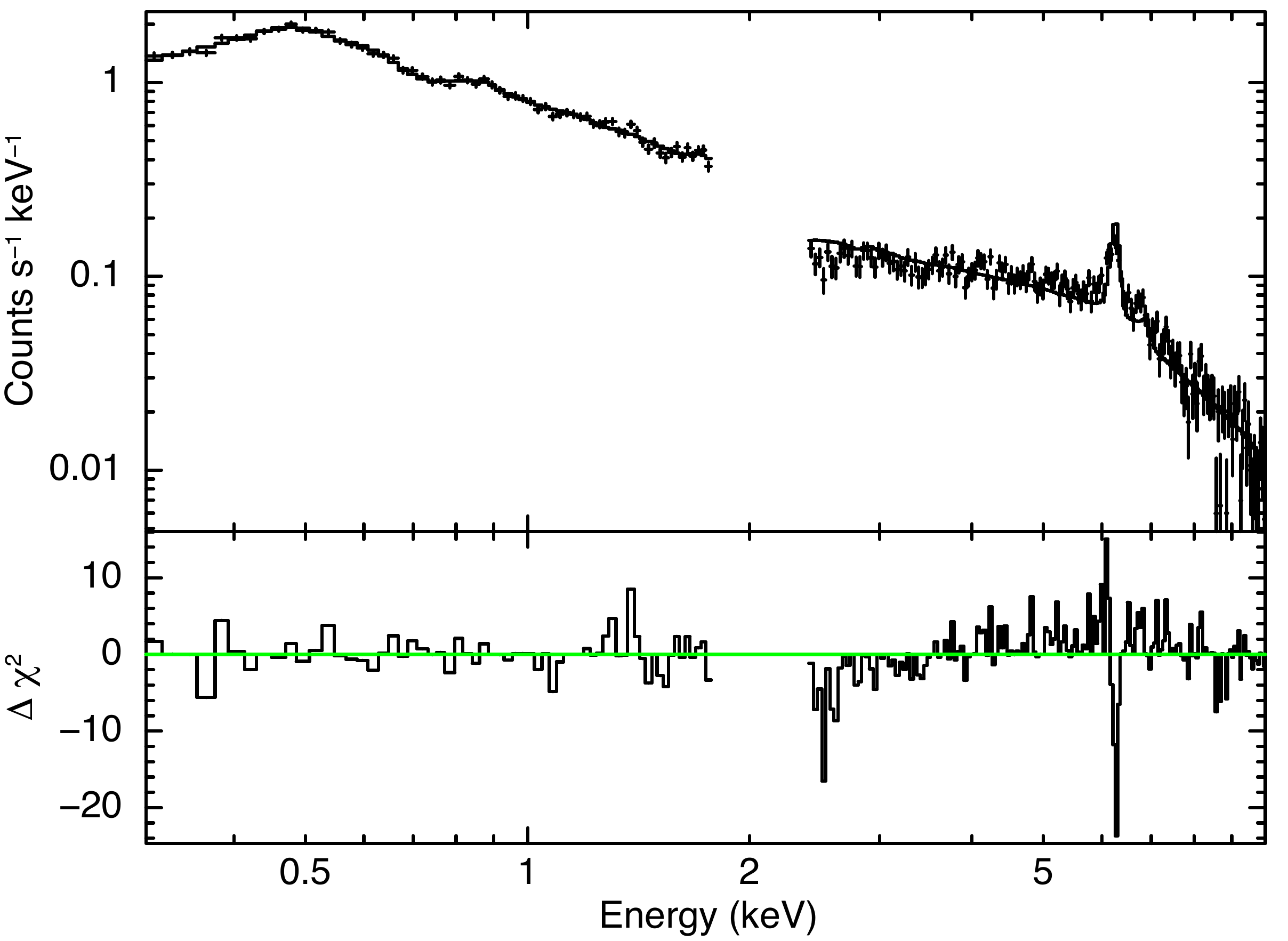}
\end{center}
 \caption{``Best fit'' for UGC 3973/ObsID:0502091001 
the worst fit obtained with our automatic fitting procedure. The presence of a feature close to the iron line 
is clearly visible.}
\label{plldara}
\end{figure}
\subsubsection{Fitting procedure}
\label{fitproc}
Given the large amount of observation data sets, we developed an automatic fitting procedure that we describe below. For each object, the main steps are:
\begin{enumerate}
\item fit all the OM-filters of all the ObsIDs of the same object with a model composed by a disk black body ({\sc diskbb} in {\sc xspec}), the ``smallBB" and the ``galaxy" template. The temperature of the disk, the ``smallBB" and ``galaxy" normalizations are let free to vary but tied between the different ObsID, while the disk normalisations are let free to vary between all the ObsIDs. We also include the reddening estimated as explained in the previous section. From the resulting best fit, we extract the normalisations of the ``smallBB" and ``galaxy", as well as the disk temperature
\item fit the OM+PN data of all the ObsIDs of the same object with the ``two-coronae'' model, fixing the disk temperature in the {\nthcomp} models  and the normalisations of the ``smallBB" and ``galaxy" components to the values obtained at the precedent step.  A first fit is done and then we let the disk temperature free to vary and refit the data again. 
\end{enumerate}
This procedure provides us with the ``two-coronae'' model best fits of all the ObsIDs of a given object. While we obtain statistically acceptable best fits for a significant number of observations, a part of the sample give quite poor results, the script converging to unrealistic parameter values, e.g. too small $\Gamma_{hc}$, of the order of unity 1, or too high disk temperature $T_{bb}>0.1$ keV. A one-by-one analysis was done then. The  too high disk temperature generally occurred when the optical-UV data of the OM filters do not enable to well constrain $T_{bb}$ and we need to fix it to its best fit value obtained in step 1 to successfully fit the OM+PN data together. Strong residuals of emission/absorption line-like features may also be present below 1 keV, and we add 2 Gaussian lines as explained in Sect. \ref{model}. In a very limited number of cases, we also add another WA component (still using a CLOUDY table). This is generally the case for objects with known complex warm absorption (see the Appendix section \ref{secapp} for discussion about these particular model fitting procedures).

\subsection{Results}
\label{res}
\subsubsection{Goodness of fits}
\label{goodfit}
The final distribution of the $\chi^2$ values we obtained as a function of the degrees of freedom (dof) of each observation is shown in Figure \ref{chi2dist}. 
The region where the adopted model can be considered correct at the 99\% confidence level is delimited by the dot-dashed curve which corresponds, for each dof value, to a 1\% probability of getting a value of $\chi^2$ as high or higher than the one we found if the model is correct. Two thirds of our fits (60\%) lie under this curve, confirming the reasonable quality of the derived spectral parameter (90\% of our sample has $\chi^2/dof<1.5$). 
Not surprisingly, most of the observations of the one third left have very large dof values due to their high SNR spectra.\\ 

The reasons of the poor fit quality are generally twofold: either a poor fit of the iron line complex and/or the presence of strong residuals in the soft X-ray part of the spectrum, where the effects of absorption from ionized matter are only roughly taken into account by our modeling. We have checked that the addition of further spectral components (WA, absorption/emission lines as explained in Sect. \ref{meth}) indeed improves the fit without modifying significantly the spectral properties of the continuum. As an example, we show in Fig. \ref{plldara} the ``best fit'' data for UGC 3973 (ObsID=0502091001) 
whose reduced $\chi^2$ (530/225) 
is among the worst of the sample. While the other data sets of UGC 3973 
give acceptable fits (see Table \ref{bestfit} in Appendix \ref{appbestfit}), this ObsID shows a clear absorption feature close to the FeK$\alpha$ line. 
The addition of a narrow ($\sigma$=0 keV) gaussian line in absorption improves drastically the fit ($\Delta\chi^2$=160 for two dof less) with a gaussian best fit energy $E_{gau}$=6.40$\pm$0.01 keV (rest frame) and EW=180 $\pm$ 30 eV. After the addition of this absorption line, the best fit parameter values for the warm and hot corona become $\Gamma_{wc}$=2.44$\pm$0.03, $kT_{wc}$=0.23$\pm$0.15 keV and $\Gamma_{hc}<$1.65. {  This absorption feature suggests a bad fit of the iron line profile. So we also test a more physical model by blurring ({\sc kdblur} model in \xspec) the reflection component given by {\sc xillver}. The inner disk radius and the inclination of {\sc kdblur} are let free to vary in the fit. The new best fit is better ($\Delta\chi^2$=103 for two dof less\footnote{The best fit parameter values of {\sc kdblurr} are $R_{in}=100^{+110}_{-40}$ and $i=19^{+3}_{-6}$}), in a smaller extend however compared to the absorption gaussian line. This could indicate an even more complex iron line profile whose precise modeling is out of the scope of the present analysis. The best fit parameter values for the warm and hot corona become $\Gamma_{wc}$=2.41$\pm$0.03, $kT_{wc}$=0.20$\pm$0.01 keV and $\Gamma_{hc}<$1.55.} These values should be compared to the ones obtained with the automatic procedure i.e. $\Gamma_{wc}$=2.40$\pm$0.03, $kT_{wc}$=0.19$\pm$0.01 keV and $\Gamma_{hc}$=1.5$_{-0}^{+0.14}$. Since the present analysis is focusing on the continuum, we safely conclude that, even when the statistical significance of the best fit is low,  the values of the X-ray parameters derived from the fit are still very reliable and can be used to properly characterize the source continuum.\\ 

This is the first important result of the application of the ``two-coronae'' model to a large sample of optical/UV/X-rays AGN spectra. The ``two-coronae'' model assumes that the optical/UV to soft X-ray emission is entirely explained by a unique warm corona component. And this model fully agrees, statistically, with the optical/UV/X-rays spectra of several AGNs without the need of an additional disk component in the optical/UV band.
\begin{figure*}
\begin{tabular}{cc}
\includegraphics[width=0.48\textwidth]{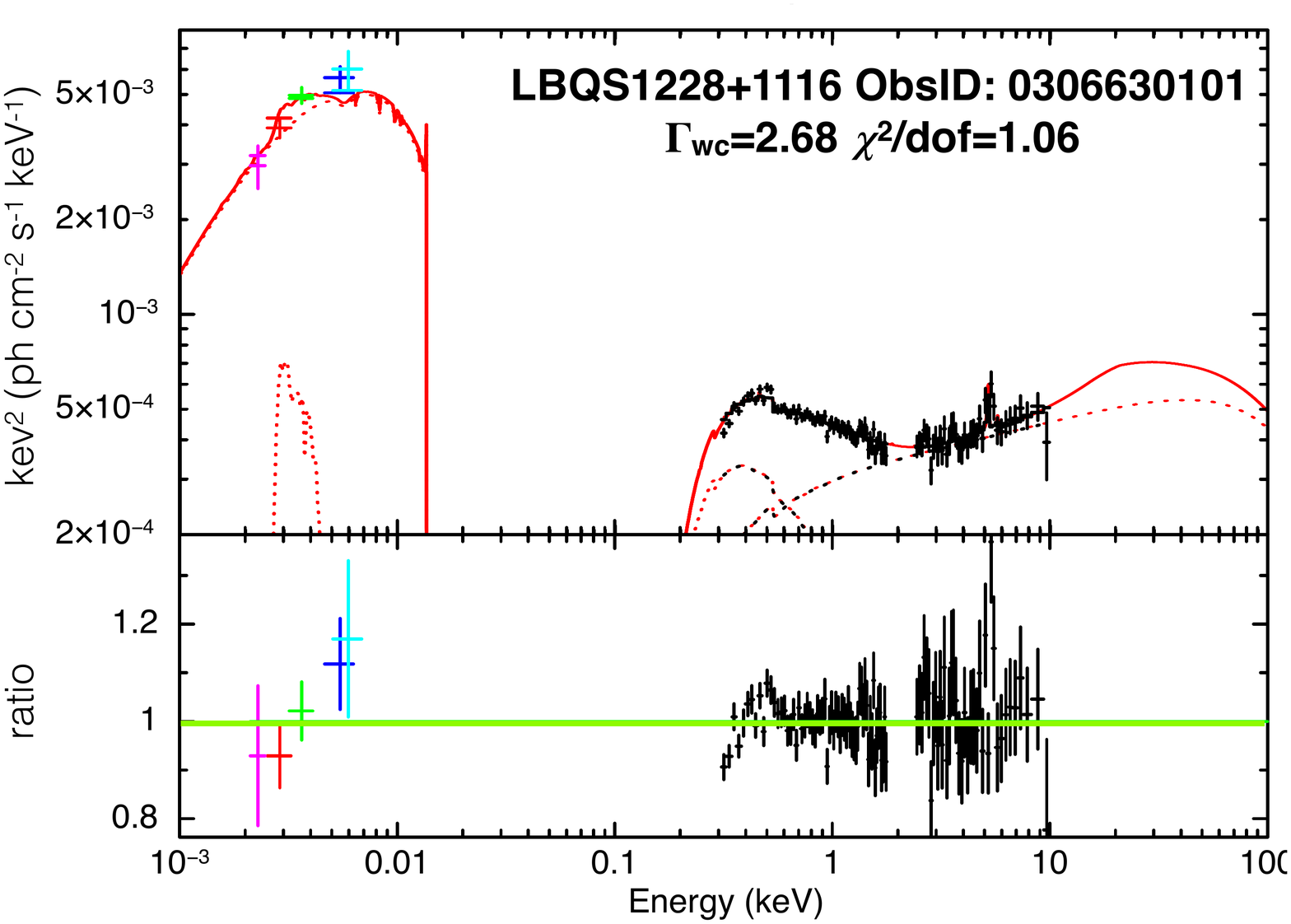}&
\includegraphics[width=0.48\textwidth]{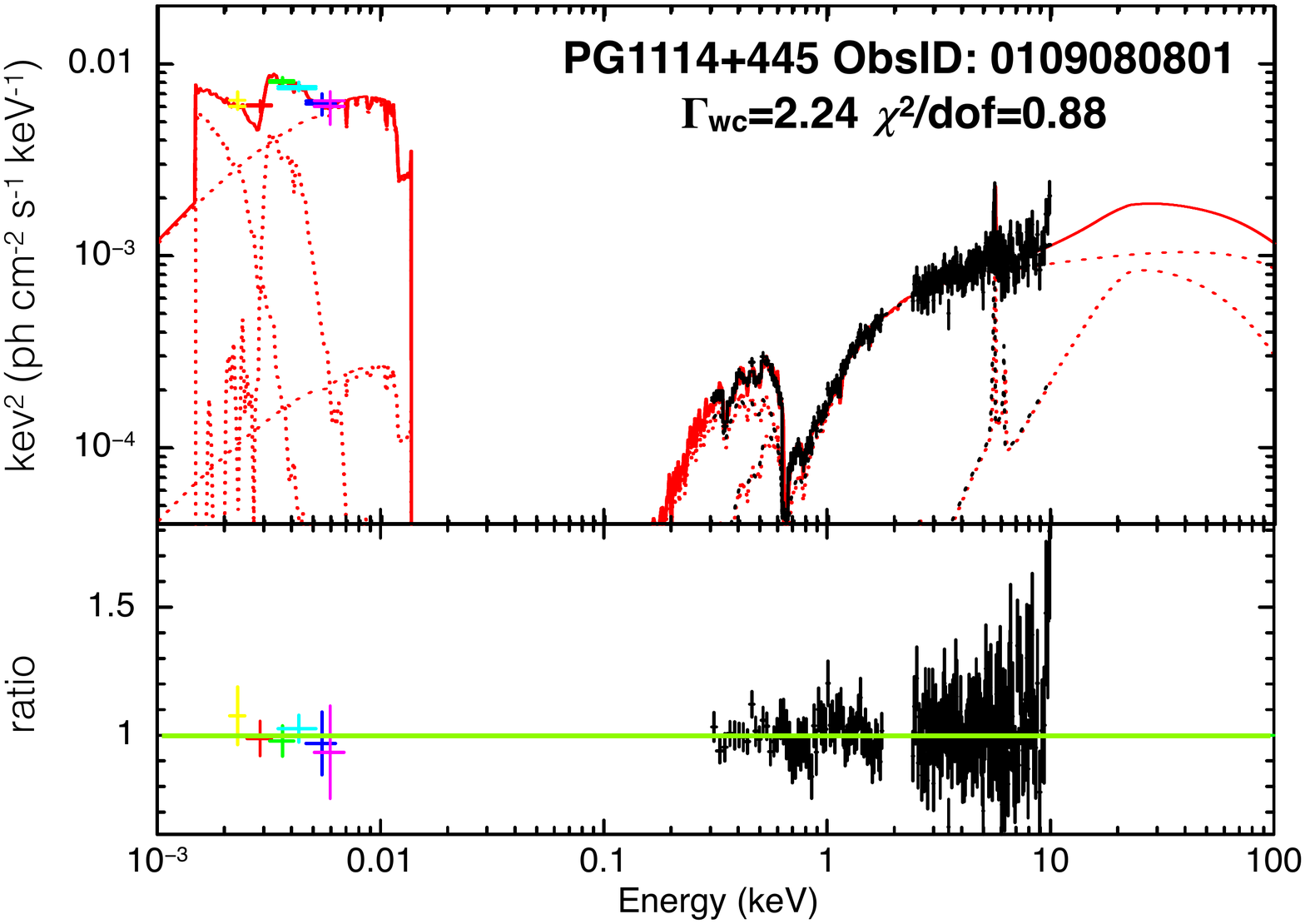}\\
\includegraphics[width=0.48\textwidth]{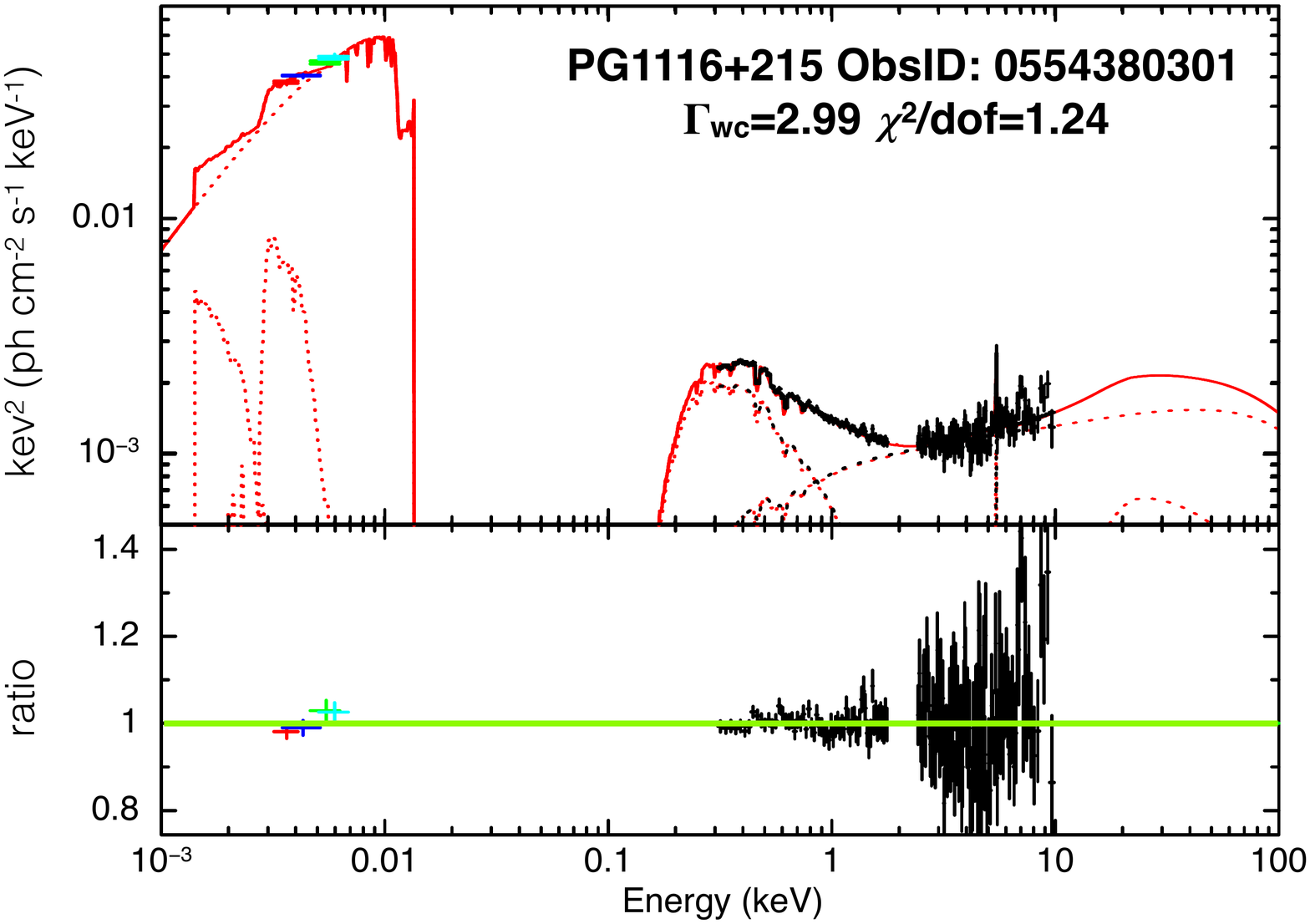}&
\begin{minipage}{0.48\textwidth}
\vspace*{-8cm} \caption{{  Examples of best fit SED and the data/model ratio for 3 observations with different warm corona photon index covering the range observed in our sample.  The source name, ObsID, $\Gamma_{wc}$ and reduce $\chi^2$ are indicated on each plot.  The black crosses represent the XMM/PN data and the color crosses the different XMM/OM filters. The red solid line is the absorbed best fit model while the dotted lines represent the different model components: hot and warm corona emission, reflection component, host galaxy  and small blue bump emission. \label{nufnu}}}
 \end{minipage}
 \end{tabular}
 \end{figure*}

{  We have reported three examples among our ``best'' fits in Fig. \ref{nufnu}. On top of each figure we have reported the OM and PN data with the unfolded best fit model as well as the corresponding data/model ratios at the bottom. These correspond to observations of three different objects with different warm corona photon indexes covering the range of $\Gamma_{wc}$ obtained in our sample, i.e., 2.24, 2.68 and 2.99 for PG1114+445,  LBPQS1228+1116 and  PG1116+215 respectively.}

\begin{figure*}
\begin{center}
\includegraphics[width=0.9\textwidth]{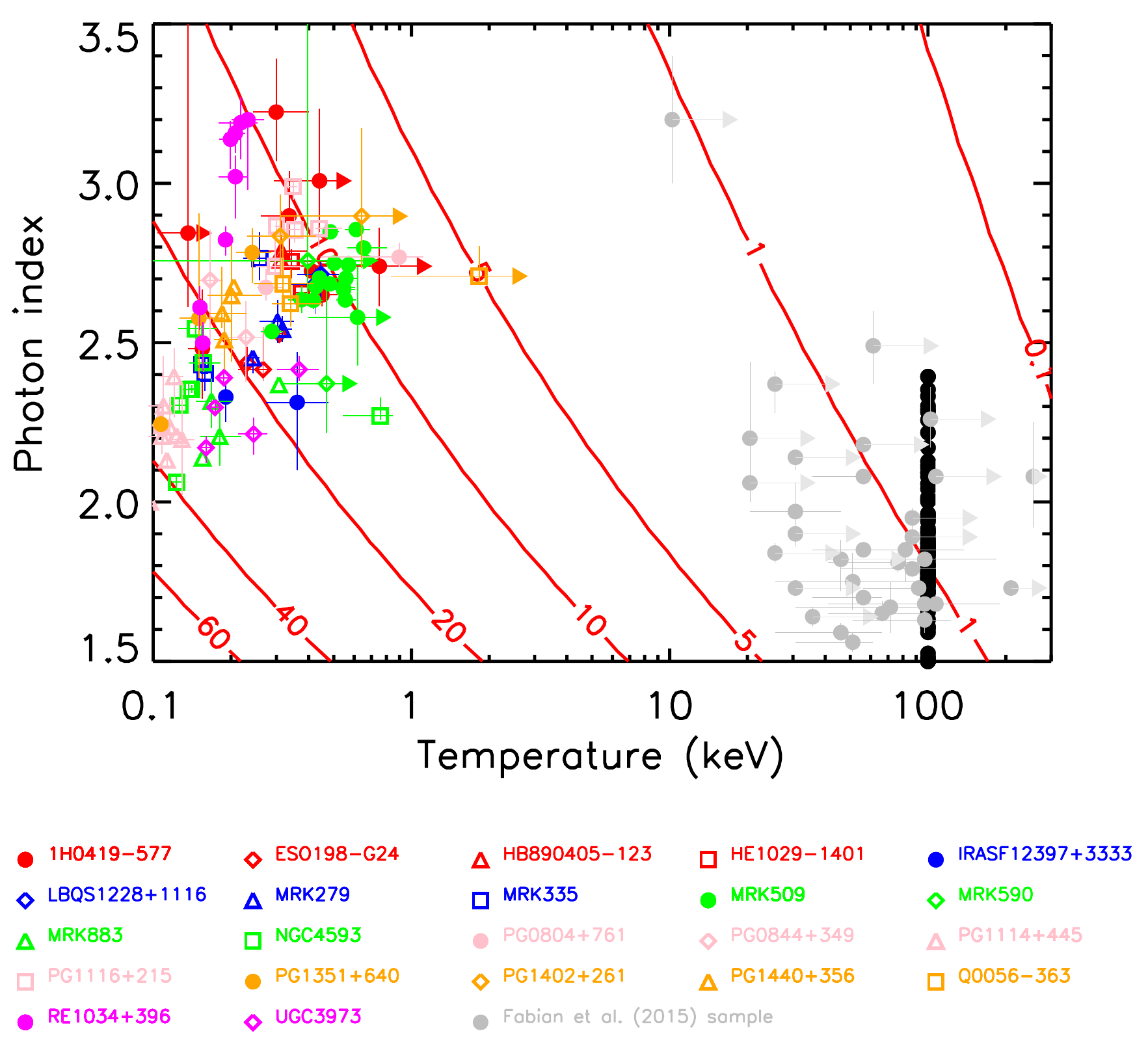}
 \caption{Best fit temperature and photon index of the warm corona of the different objects of our sample. Different colored symbols are used for each object. They cluster in the left part of the figure, and agree with large ($>$5) optical depths. For comparison, we have also reported the best fit photon indexes of the hot corona (black circles on the right). The temperature of the hot corona being fixed to 100 keV in the fits, they are all aligned vertically. The gray filled circles correspond to the best fit parameters of the hot corona from \cite {fab15}. The hot corona parameters cluster in the bottom right part of the figure where the optical depth is close to unity.
 }
\label{figgamkt_obs}
\end{center}
\end{figure*}

\begin{figure}
\begin{tabular}{c}
\includegraphics[height=6cm]{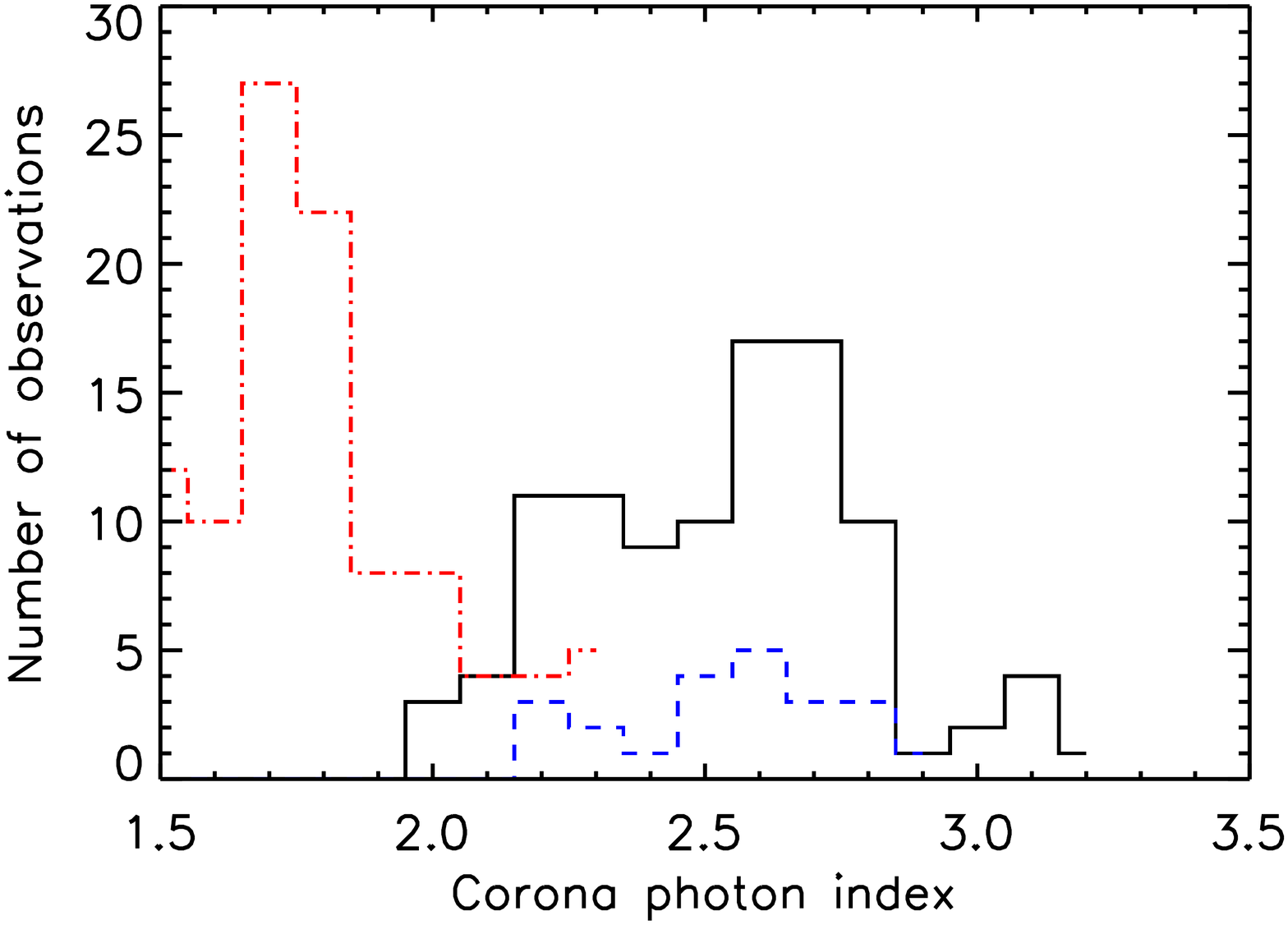}\\
\includegraphics[height=6.25cm]{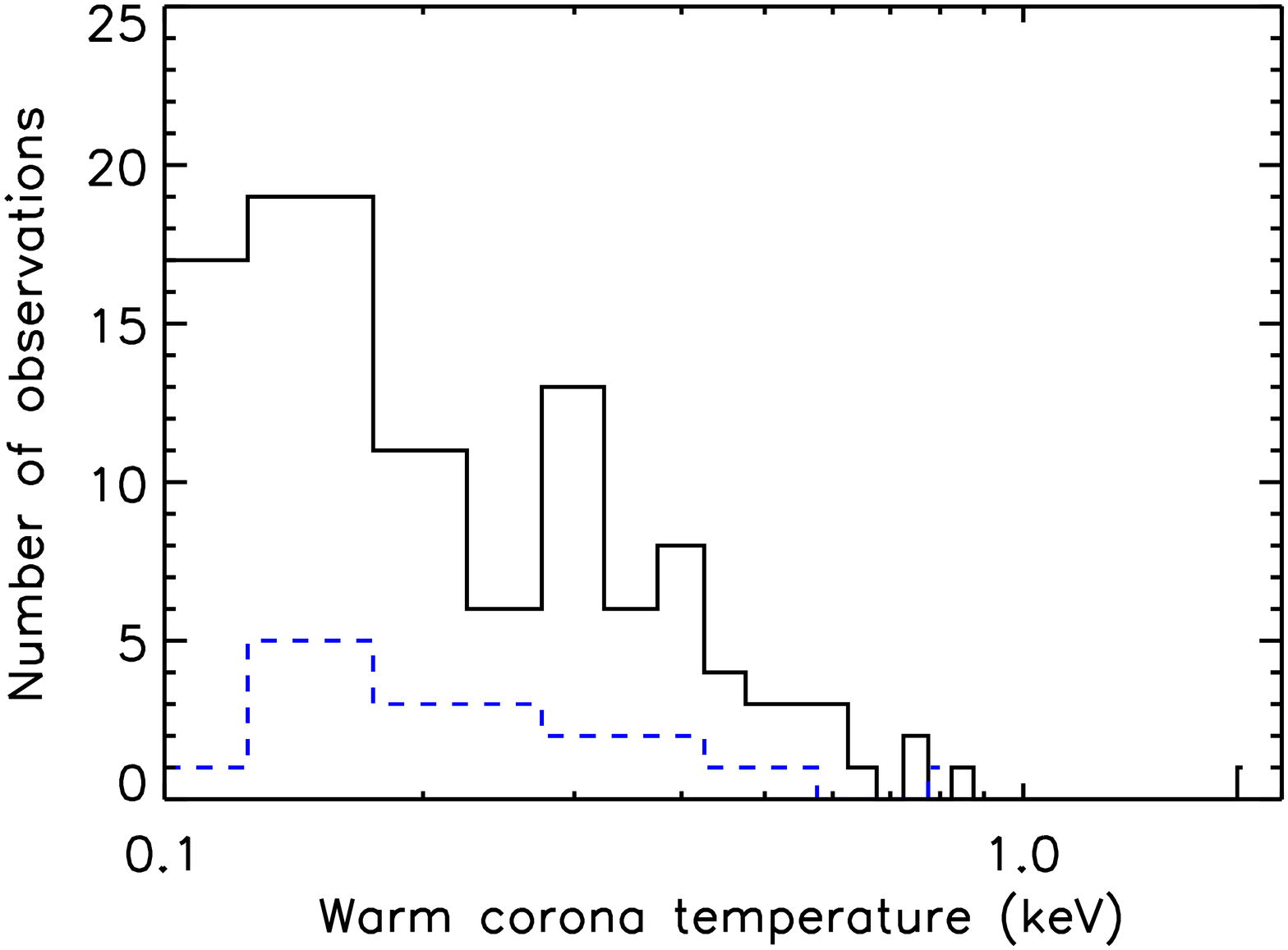}
\end{tabular}
 \caption{Histogram of the warm corona photon index $\Gamma_{wc}$ (top) and temperature $kT_{wc}$ (bottom) of the warm corona model for the different objects of our sample. The black solid histograms correspond to the whole observations.The dashed (blue) histogram correspond to the average values of $\Gamma_{wc}$ and $kT_{wc}$ for each object. On the top figure, we have also overplotted in dot-dashed (red) line, the histogram of   the hot corona photon index $\Gamma_{hc}$.}
\label{figgamkthisto}
\end{figure}

\subsubsection{The warm corona physical properties}
The best fit parameter values of the warm corona (i.e. photon index $\Gamma_{wc}$ and temperature $kT_{e,wc}$) as well as the hot corona photon index $\Gamma_{hc}$, the reduced $\chi^2$ and the degrees of freedom are reported in Table \ref{bestfit} of Appendix \ref{appbestfit}. The warm corona photon index and temperature are also reported in Fig. \ref{figgamkt_obs} for all the sources of our sample  and the histograms of their distribution are shown in Fig. \ref{figgamkthisto}. The optical depth contour plots (from Fig. \ref{figgamkt}) are overplotted in Fig. \ref{figgamkt_obs}  to ease the comparison. 

The $\Gamma_{wc}$ values are distributed between 2 and 3, with a clear preferential value around 2.6-2.7. The warm corona temperature $kT_{e,wc}$ is in the 0.1-1 keV range, but it is preferentially smaller than 0.4 keV. This corresponds to a region of the $\Gamma$-$kT$ plane  with large optical depths, most of the sources agreeing with $\tau_{warm}>10$. This is in agreement with similar spectral analysis published in the literature (e.g. \citealt{pet13,jin12c}). The dashed (blue) histograms in Fig. \ref{figgamkthisto} correspond to the histograms of the averaged values of $\Gamma_{wc}$ and  $kT_{e,wc}$ for each objects of our sample. They show roughly the same trends as those observed for the whole observation sample.\\

For comparison we have overplotted in Fig. \ref{figgamkt_obs}  the best fit photon indexes of the hot coronae $\Gamma_{hc}$ (black circles on the right). The temperature of the hot corona being fixed to 100 keV in the fits, they are all aligned vertically. The histogram of the $\Gamma_{hc}$ values is overplotted in Fig. \ref{figgamkthisto}. It peaks between 1.5 and 2. We have also reported in Fig. \ref{figgamkt_obs}, with gray filled circles, the best fit parameters of the hot coronae from \cite {fab15}.  The warm and hot coronae parameters a clearly separated, the hot coronae parameters clustering at the bottom right part of the figure with harder spectra ($\Gamma<$1.9) and higher temperature ($kT>$20 keV) characteristic of the high energy emission of type 1 objects. The hot corona optical depths are also of the order of unity as generally observed (e.g. \citealt{pet01,mat14,bren14a,bren14b}).

\subsubsection{A slab-like geometry above a passive accretion disk}
\begin{figure}
\includegraphics[width=0.95\columnwidth]{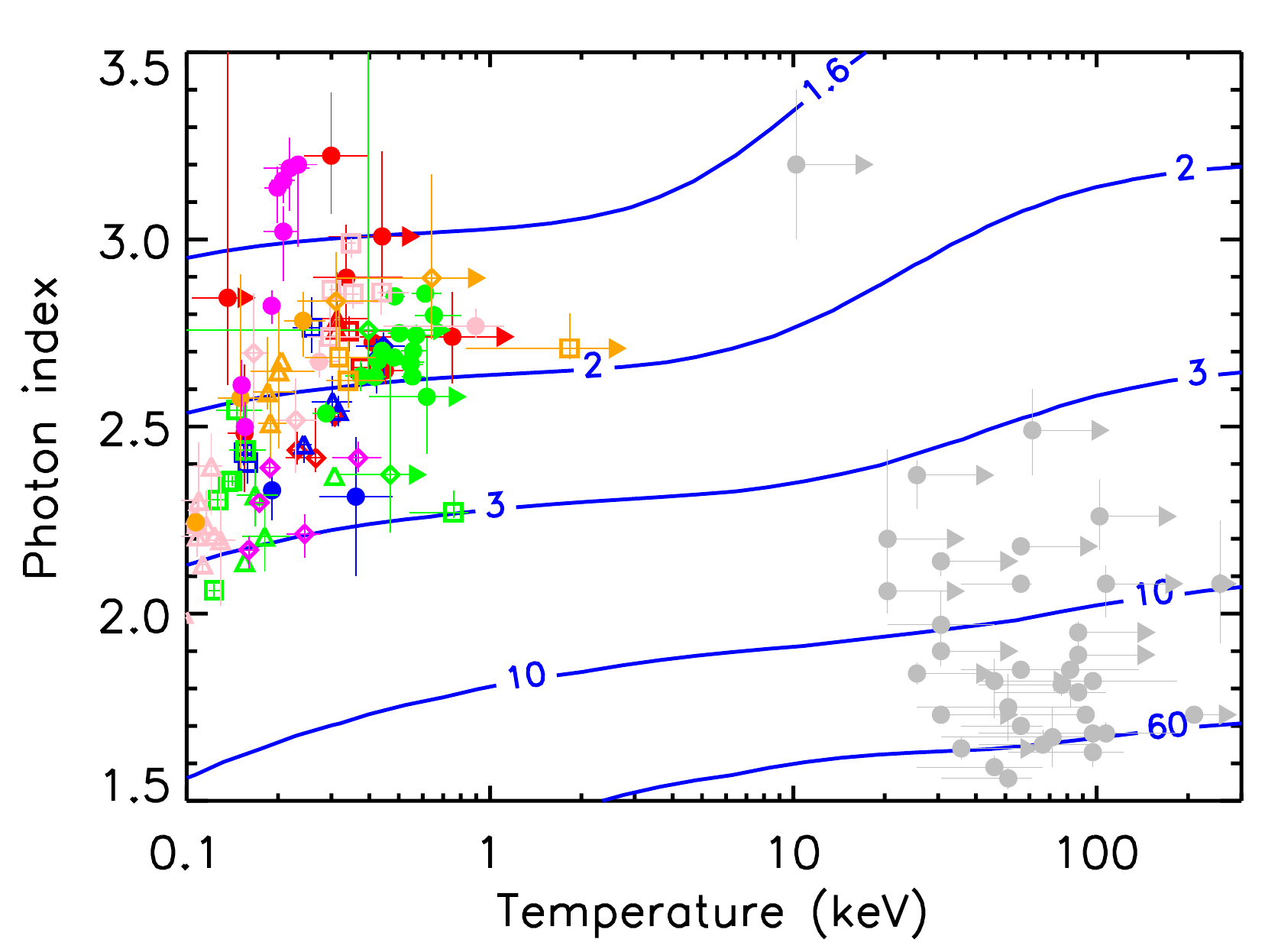}
 \caption{Contours of the amplification factor $A$ in the corona spectral parameter $\Gamma$-$kT$ plane. The colored symbols and the gray filled circles correspond to the same legends as in Fig. \ref{figgamkt} 
\label{figgamktb} }
\end{figure}

\begin{figure}
\includegraphics[width=0.95\columnwidth]{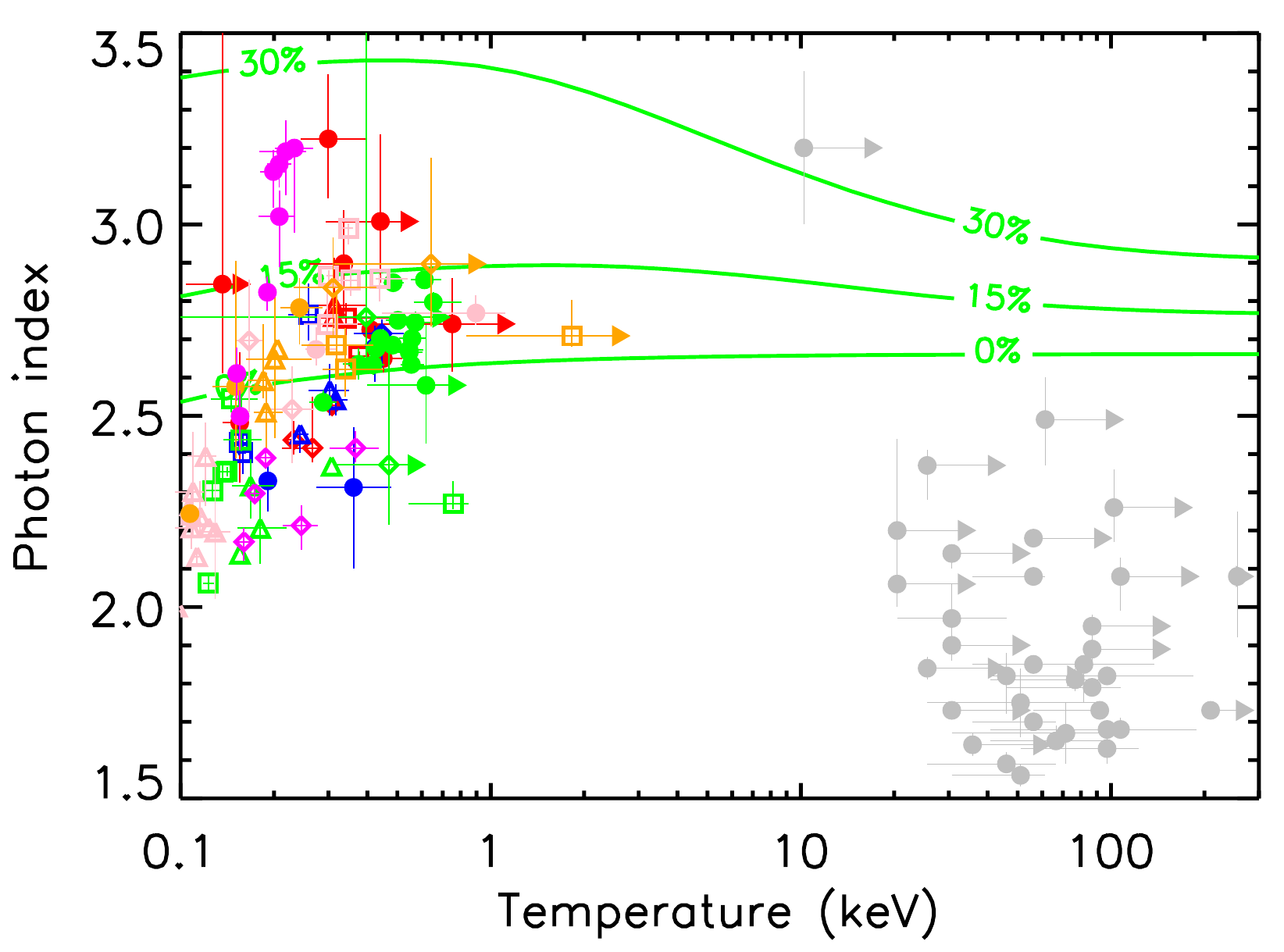}
 \caption{Contours of the minimal fraction (in \%) of disk intrinsic  emission $\displaystyle \left . \frac{L_{disk,intr}}{L_s}\right |_{min}$ in the corona spectral parameter $\Gamma$-$kT$ plane. The colored symbols and the gray filled circles correspond to the same legends as in Fig. \ref{figgamkt} 
\label{figgamktc} }
\end{figure}

We have reported our best fit parameter values for the warm corona in Fig. \ref{figgamktb} and \ref{figgamktc} where the contours of, respectively, the amplification factor and $\displaystyle \left . \frac{L_{disk,intr}}{L_s}\right |_{min}$ (from Fig. \ref{figgamkt}) have been overplotted. The warm corona amplification ratios are distributed between 1.6 and 3, with a peak (corresponding to the peak of the photon index around 2.6-2.7 in Fig. \ref{figgamkthisto}) close to 2. This is precisely the theoretical value of a thick corona in slab geometry and in radiative equilibrium with a passive disk as discussed in Sect. \ref{maineq}. 

This is also shown in Fig. \ref{figgamktc}, since all the objects agree with an intrinsic disk emission lower than $\sim$20\%, i.e. the disk is mostly passive, radiating only through the reprocessing of the warm corona emission.  As already suggested in the dedicated analysis of Mkn 509 (P13), a warm corona above a passive disk appears as a reasonable (from a statistical point of view) explanation of the soft X-ray excess in AGN.\\

About half of our sample however are below the contour  $\displaystyle \left . \frac{L_{disk,intr}}{L_s}\right |_{min}$=0\% with an amplification ratio larger than 2, in between 2 and 3. As discussed in Sect. \ref{maineq}, this suggests a patchy warm corona. In the case of a corona above the accretion disk,  we could use Eq. \ref{eqratio} to give some contraints on the corona ``patchiness". Indeed we have:
\begin{equation}
g=\frac{2}{A-e^{-\tau}+2\frac{L_{disk,intr}}{L_s}}.
\end{equation}
Assuming a passive disk (and large $\tau$ as estimated for the warm corona), we obtain $2/3<g<1$ for $A$ between 2 and 3. In term of solid angle sustained by the warm corona, this translates to $4\pi/3<\Omega_{wc}<2\pi$, i.e., a slightly patchy corona with respect to the slab.

By comparison, and as commonly observed, the hot corona parameters agree with large amplification ratios (mainly above 10), signature of a very photon-starved geometry. Using again the above expression (still assuming a passive disk but now with a corona optical depth $\tau=1$), we indeed find $g<0.2$ for $A>10$, or $\Omega_{hc}<2\pi/5$.\\

While crude and limited to the ``two-coronae'' framework, these estimates of the warm and hot coronae ``patchiness" support a disk-coronae geometry where the hot corona is localized in the inner part of the accretion flow (where the hottest temperatures are expected to be reached) while the warm corona largely covers the outer part of an optically thick accretion disk (see e.g. Fig. 10 of P13). The ``lampost''  geometry, where the hot corona lies above the black hole and illuminates an accretion disk covered by a warm corona is also consistent with these estimates. This is however different from the geometries proposed by e.g. \cite{don12} in which, by construction, the accretion disk is intrinsically luminous, which implies a smaller solid angle sustained by the warm corona.\\ 

\subsubsection{Correlation}
\begin{figure}
\includegraphics[width=\columnwidth]{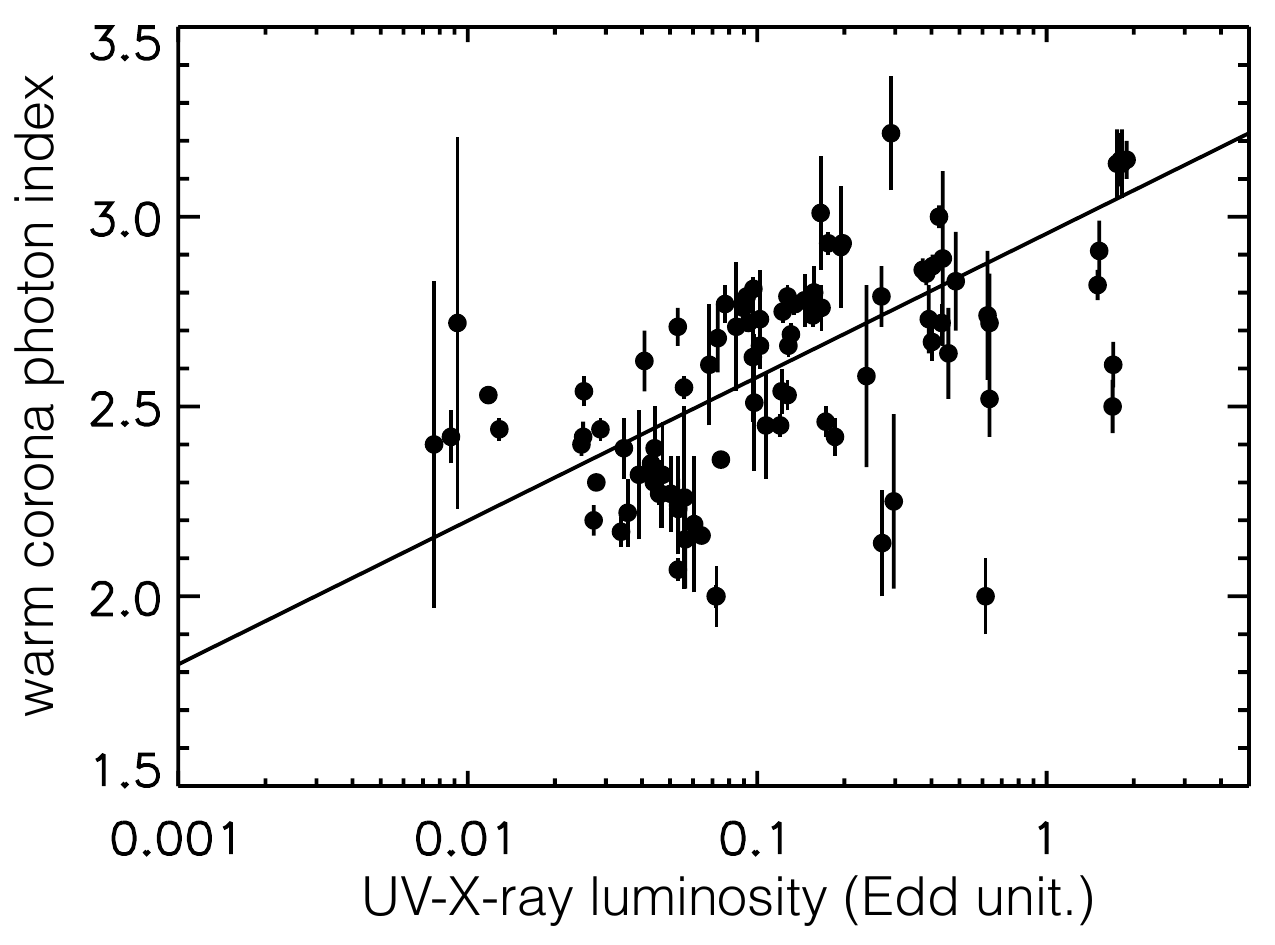}
 \caption{Plots of the warm corona photon index versus the UV/X-ray luminosity (in Edd. unit) for all the observations of our sample but  those of HB890405-123 and LBQS1228+1116 for which we do not have black hole mass estimates. The solid line corresponds to the log-linear best fit. \label{uncorr}.}
 \end{figure}

{  We do not find any significant correlations between the different best fit spectral parameters. We check also for correlations with the black hole mass and the UV/X-ray luminosity $L_{UV-X}$. We estimate the latter by integrating the flux of our unabsorbed SEDs. Using the estimate of the black hole mass for a large part of our sample (see Tab. \ref{tab1} and \cite{bia09a} for the references. Only HB890405-123 and LBQS1228+1116 have no BH mass estimates) we can then estimate 
$L_{UV-X}/L_{Edd}$. These values have been reported in Tab. \ref{bestfitlbol}. Interestingly, a strong correlation (with a linear Pearson correlation coefficient equal to 0.47, which corresponds to a $>$99\% confidence level for its significance) 
is found between $\Gamma_{wc}$ and $L_{UV-X}/L_{Edd}$. It is reported in Fig. \ref{uncorr}. A log-linear fit gives:
\begin{equation}
\Gamma_{wc}=(2.95\pm0.01)+(0.38\pm0.01)\log\left (\frac{L_{UV-X}}{L_{Edd}}\right )
\end{equation}
and is overplotted in the figure with a solid line. The slope is quite small and indicate a rather smooth dependency. The standard deviation of all the points with respect to the best fit relation is $\sim$0.2, slightly smaller than the dispersion of the warm corona photon index ($\sim$0.3). Anyway, since large $\Gamma_{wc}$($>$2.6) agree, in our two-coronae model, with larger intrinsic disk emission, this correlation suggests that high accretion rate systems would go with more intrinsic disk dissipation. }


\section{Concluding remarks}
In this study we discuss the theoretical expectations of the spectral parameters (i.e. photon index $\Gamma$ and temperature $kT_e$) of the comptonisation emission of a thermal corona in radiative equilibrium with the accretion disc. We detail how the corona optical depth $\tau$, its amplification factor $A$ and the minimal intrinsic disk emission  $\displaystyle \left . \frac{L_{disk,intr}}{L_s}\right |_{min}$ can be mapped in the $\Gamma$-$kT_e$ plane. \\

Then, we show that the spectral constraints published in the literature for the soft X-ray excess suggest an optically thick ($\tau\sim$ 10-20), extended (amplification factor $A\simeq$2) thermal corona in radiative equilibrium above a weakly dissipative disc. On the contrary, and as a well known result, the hard X-ray emission agrees with an optically thin ($\tau\sim$ 1) and patchy ($A\gg$1) thermal corona. \\

To better test this ``two-coronae'' dichotomy, we apply the ``two-coronae'' model to a large sample of AGN with simultaneous optical/UV and X-ray data from XMM-Newton. Our sample is composed by 22 AGN and 100 ObsID. In the ``two-coronae'' model, the warm corona reproduces the entire optical/UV and soft X-ray emission while the hot corona explains the high energy ($>$ 2 keV) emission.  

The ``two-coronae'' model gives a statistically reasonable fit to the optical/UV/X-ray data of our sample.  Our best fit physical parameters for the warm corona indicate an optical depth $\tau\sim$10-40 and an electron temperature $kT\sim$0.1-1 keV. In comparison, the hot corona is optically thin ($\tau\sim$1) and hot (kT$>$20 keV). More interestingly, we confirm that the warm corona parameters agree well with a quite extended corona above a passive disk, i.e. most of the accretion power would be released in the warm corona while the disk would only reradiate the emission coming from the corona. This result was firstly suggested by P13 who apply the ``two-coronae'' model to the monitoring of Mkn 509. The present study shows that this interpretation could apply to a large number of objects suggesting a geometry similar to Fig. 10 of P13 for the inner accretion flow. \\

{  If this interpretation seems qualitatively correct, the observed dispersion, in the sample, of the warm corona parameter values may indicate a true dispersion in the intrinsic disc and disc/corona geometry properties. Some objects like, e.g., RE1034+396, show a quite steep warm corona photon index $\sim$3 suggesting a more active underlying disc than the other sources, consistent with the known high $L/L_{Edd}$ in this object (see Tab. \ref{bestfitlbol} but also \citealt{jin12a,don12}). On the contrary, other objects like, e.g., PG1114+445, have a quite hard warm corona photon index $\sim$2.2 producing a weak "bend" (and consequently a weak soft X-ray excess) with respect to the hard X-ray power law. Alternatives for the soft X-rat excess, as for example blurred reflection and/or absorption, could play a significant role here (see similar discussion in \citealt{don12,jin12a}). A more careful modelling, out of the scope of the present paper, would be required to test these hypotheses. }

{  Our modeling of the hot and warm coronae with a one-temperature plasma is also certainly crude. A temperature distribution is expected, globally decreasing with radius. Thus, if the warm corona could be radially elongated, as suggested by our results, it is probably not covering the entire disk and part of the observed optical emission could come from the outer part of the disk. Optical/UV reverberation studies also suggest that the accretion flow geometry could be quite complex and that the optical/UV emission could be produced by more than a single component (e.g. \citealt{gar16} and references therein). The addition of a model component in the optical in our study would naturally imply a less extended warm corona, more localized in the inner (UV/FUV emitting) region of the accretion flow, and ,consequently, a larger intrinsic disk emission to explain the observed values of warm corona temperature and photon index. While we do not expect this effect to significantly change our conclusion, its precise estimate would also require a detailed modeling of the optical/UV emission  out of the scope of the present paper given the spectral imitation of the use of the XMM/OM large band filters.}\\

Some radio-quiet AGN show clear signatures of relativistic iron lines, while the absence of such component in the others could be due to a lack of statistics (e.g. \citealt{bren09,delacal10,man16}). Relativistically blurred ionized reflection can produce significant emission in the soft X-rays and could even explain a large part of the soft X-ray excess (e.g. \citealt{cru06}). Due to the additional complexity, our fits do not include a relativistic reflection. Our reduced $\chi^2$ are also reasonable good and do not require the addition of such component. Now, if it was present, it would produce soft X-ray emission and, consequently, it is expected to weaken the amount of soft X-ray emission coming from the warm corona. The fits would compensate with a larger values of $\Gamma_{wc}$, moving our sample to the top of Fig. \ref{figgamkt} i.e. the minimal fraction of the disk intrinsic emission would increase with respect to the present estimates. We do not expect, however, drastic changes of $\Gamma_{wc}$. And we believe that the data would still agree with a quite low fraction of disk intrinsic emission ($<$50\%), most of the accretion power being still released in the warm corona.\\

The presence of such a corona above the accretion disk may have important consequences, with direct impact on e.g. our understanding of the accretion disk vertical equilibrium, the expected spectral emission from such accretion flow or its capacity of producing outflows/jets, etc...  Moreover it would certainly have some impacts on the spectral shape of the reflection component. In the case of a hot ($kT\simeq$100 keV) and optically thin ($\tau<$1) corona, the reflection component is expected to be partly smeared out when crossing the corona, modifying the shape of the reflection hump and the iron line equivalent width measurements (e.g. \citealt{pet01,stei17}). Similar effects are expected in the warm corona. The cooler temperature will rather produce the down scattering of the illuminating photons (e.g. the iron line should be redshifted) but some broadening should be present too. On the other hand, the warm corona temperature of the order of 0.5-1 keV is such that several ionized lines should be produced and potentially observed. These are important signatures that has to be tested with accurate radiative transfer codes in order to confirm or rule out the presence of such a warm plasma in the inner part of the accretion flow.\\


 
\section*{Acknowledgments}
We would like to thank the referee for her/his careful reading of the manuscript and useful comments that help to improve its quality. This work is based on observations obtained with XMM-Newton, an ESA science mission with instruments and contributions directly funded by ESA Member States and the USA (NASA). We acknowledge financial support from the CNRS/INAF french/Italian PICS programme. POP acknowledges financial support from CNES and the French PNHE. POP and FU acknowledge support from the Italo/French Vinci programme. The research leading to these results has received funding from the European Union's Horizon 2020 Programme under AHEAD project (grant agreement n. 654215). SB and GM acknowledge financial support from the European Union Seventh Framework Programme (FP7/2007-2013) under grant agreement no. 31278. MD and MC acknowledges support from the ASI - INAF grant I/037/12/0


\begin{appendix}
{  
\section{The spectral index of the compton emission from a slab and optically thick corona above a passive disk}
\label{app1}
We demonstrate here that the constraints on the photon rate and the amplification factor (see Sect. \ref{maineq}) impose the corona photon index to be close to 2.6. For simplification, let us assume the corona to be spherical, with radius $R$, and the disk emission spectral density to have a black body distribution (instead of a disk black body distribution), characterized by a temperature $T_{bb}$, homogeneously distributed inside the corona. This is different from a slab geometry where the disk emission comes from below the corona. For a spherical, optically thick corona, Eqs. \ref{dotn} and \ref{eqAb}  becomes:
 \begin{eqnarray}
\dot{n}_s\simeq \dot{n}_{obs}\label{eqfirst}\\
L_s\simeq L_{obs}.
\end{eqnarray}
Then, from the integration of the black body distribution, the disk photon rate $\dot{n}_s$ and luminosity $L_{s}$ are given by:
\begin{eqnarray}
\dot{n}_s &=& \frac{16\pi\zeta(3)k^3}{h^3 c^3} T_{bb}^3 4\pi R^2c\\
 &=& 2.029 \times 10^7 T_{bb}^3 4\pi R^2c \mbox{\hspace{0.5cm} ph.s$^{-1}$}\\
{L}_s &=& \frac{8\pi^5 k^4}{15h^3 c^3} T_{bb}^4 4\pi R^2c \\
 &=&7.6 \times 10^{-16} T_{bb}^4 4\pi R^2c \mbox{\hspace{0.5cm} J.s$^{-1}$}
\end{eqnarray}
with $\zeta(3)$, the Apery constant (equal to $\sim$1.2). 

On the other hand, approximating the Compton corona spectrum by a simple power law, $F_E=N_0 E^{-\Gamma}$, between $E_{bb}=kT_{bb}$ and a cut-off energy $E_{c}$, the observed corona photon rate $\dot{n}_{obs}$ and luminosity $L_{obs}$ are given by:
\begin{eqnarray}
\dot{n}_{obs} &=& \frac{N_0}{1-\Gamma} (E_{c}^{1-\Gamma}-E_{bb}^{1-\Gamma})\\
{L}_{obs} &=& \frac{N_0}{2-\Gamma} (E_{c}^{2-\Gamma}-E_{bb}^{2-\Gamma}).\label{eqlast}
\end{eqnarray}
Combining these different equations (from \ref{eqfirst} to \ref{eqlast}) and assuming $E_c\gg E_{bb}$ (i.e. neglecting $E_{c}^{1-\Gamma}$ and $E_{c}^{2-\Gamma}$ with respect to $E_{bb}^{1-\Gamma}$ and $E_{bb}^{1-\Gamma}$ respectively) we obtain a simple relation on $\Gamma$:
\begin{equation}
\frac{1-\Gamma}{2-\Gamma}\simeq 2.7 \mbox{          i.e.             }\Gamma\simeq 2.6.
\end{equation}

}
\section{The sample}

%
%
{
\begin{longtable}{cccccccc}
\hline\hline
Source  & RA & Dec & Redshift & $\log M_{bh}$ & E(B-V)& ObsID & Available OM filters \\
\hline
\endfirsthead
\caption{continued.}\\
\hline\hline
Source  & RA & Dec & Redshift & $\log M_{bh}$ & E(B-V)& ObsID & Available OM filters \\
\hline
\endhead
\hline
\endfoot
1H0419-577  &  66.5030 & -57.2002 & 0.1040 & 8.580 & 0.0296 &  0148000201  &  U, B, V, W1, W2\\
 & & & & & &  0148000301  &  U, B, V, W1, W2\\
 & & & & & &  0148000401  &  U, B, V, W1, W2\\
 & & & & & &  0148000501  &  U, B, V, W1, W2\\
 & & & & & &  0148000601  &  U, B, V, W1, W2\\
 & & & & & &  0604720301  &  B, W1, M2, W2\\
 & & & & & &  0604720401  &  B, W1, M2, W2\\
 & & & & & &  0112600401  &  W1, \\
ESO198-G24  &  39.5818 &  -52.1923 &  0.0455 &  8.48 &  0.0456 &  0305370101  &  U, B, W1, M2, W2\\
 & & & & & &  0112910101  &  U, \\
 & & & & & &  0067190101  &  W1, W2\\
HB890405-123  &  61.9517 &  -12.1935 &  0.5725 &  - &  0.0556 &  0202210401  &  U, W1, W2\\
 & & & & & &  0202210301  &  U, W1, W2\\
HE1029-1401  &  157.976 &  -14.2808 &  0.0858 &  8.73 &  0.0954 &  0203770101  &  U, V, W1, M2, W2\\
 & & & & & &  0110950101  &  W2\\
IRASF12397+3333  &  190.544 &  33.2840 &  0.0435 &  6.66 &  0.0194 &  0202180301  &  U, B, W1, M2, \\
 & & & & & &  0202180201  &  U, B, V, W1, M2, W2\\
LBQS1228+1116  &  187.725 &  11.0031 &  0.2362 &  - &  0.0313 &  0306630201  &  U, B, V, W1, M2, W2\\
 & & & & & &  0306630101  &  U, B, V, M2, W2\\
MRK279  &  208.264 &  69.3082 &  0.0304 &  7.54 &  0.0259 &  0302480501  &  U, W1, M2, W2\\
 & & & & & &  0302480601  &  U, W1, M2, W2\\
 & & & & & &  0302480401  &  U, W1, M2, W2\\
MRK335  &  1.58130 &  20.2029 &  0.0257 &  7.15 &  0.0582 &  0600540601  &  U, B, V, W1, M2, W2\\
 & & & & & &  0600540501  &  U, B, V, W1, M2, W2\\
 & & & & & &  0510010701  &  U, B, W1, M2, W2\\
MRK509  &  311.0405 & -10.7234 & 0.0343 & 8.16 & 0.0600 &  0601390201  &  U, B, V, W1, M2, W2\\
 & & & & & &  0601390301  &  U, B, V, W1, M2, W2\\
 & & & & & &  0601390401  &  U, B, V, W1, M2, W2\\
 & & & & & &  0601390501  &  U, B, V, W1, M2, W2\\
 & & & & & &  0601390601  &  U, B, V, W1, M2, W2\\
 & & & & & &  0601390701  &  U, B, V, W1, M2, W2\\
 & & & & & &  0601390801  &  U, B, V, W1, M2, W2\\
 & & & & & &  0601390901  &  U, B, V, W1, M2, W2\\
 & & & & & &  0601391001  &  U, B, V, W1, M2, W2\\
 & & & & & &  0601391101  &  U, B, V, W1, M2, W2\\
 & & & & & &  0306090101  &  W1, \\
 & & & & & &  0306090201  &  M2, W2\\
 & & & & & &  0306090301  &  W1, M2, W2\\
 & & & & & &  0306090401  &  W1, M2, W2\\
 & & & & & &  0130720101  &  M2, W2\\
 & & & & & &  0130720201  &  U, W1, W2\\
MRK590  &  33.6398 &  -0.766600 &  0.0263 &  7.68 &  0.0391 &  0109130301  &  U, B, V, W1, M2, W2\\
 & & & & & &  0201020201  &  U, B, W1, M2, W2\\
MRK883  &  247.470 &  24.4439 &  0.0374 &  7.28 &  0.0579 &  0652550201  &  U, B, W1, \\
 & & & & & &  0302260701  &  U, B, W1, M2, W2\\
 & & & & & &  0302261001  &  U, B, W1, M2, W2\\
 & & & & & &  0302260101  &  U, B, W1, M2, W2\\
NGC4593  & 189.9142 & -5.3442 & 0.0090 & 6.729 & 0.0337  &  0740920201  &  U, W1, W2\\
 & & & & & &  0740920301  &  U, W1, W2\\
 & & & & & &  0740920401  &  U, W1, W2\\
 & & & & & &  0740920501  &  U, W1, W2\\
 & & & & & &  0740920601  &  U, W1, W2\\
 & & & & & &  0109970101  &  W1, W2\\
 & & & & & &  0059830101  &  W1, M2, W2\\
PG0804+761  &  122.744 &  76.0451 &  0.1000 &  8.24 &  0.0435 &  0605110101  &  U, V, W1, M2, \\
 & & & & & &  0605110201  &  U, V, W1, M2, \\
PG0844+349  &  131.927 &  34.7512 &  0.0640 &  7.97 &  0.0480 &  0554710101  &  U, B, V, W1, M2, W2\\
 & & & & & &  0103660201  &  U, \\
PG1114+445  &  169.277 &  44.2259 &  0.1438 &  8.59 &  0.0264 &  0651330201  &  W1, M2, W2\\
 & & & & & &  0651330101  &  W1, M2, W2\\
 & & & & & &  0651330301  &  W1, M2, W2\\
 & & & & & &  0651330401  &  W1, M2, W2\\
 & & & & & &  0651330601  &  W1, M2, W2\\
 & & & & & &  0651330901  &  W1, M2, W2\\
 & & & & & &  0651330801  &  W1, M2, W2\\
 & & & & & &  0651331001  &  W1, M2, W2\\
 & & & & & &  0651330501  &  W1, M2, W2\\
 & & & & & &  0109080801  &  U, B, V, W1, M2, W2\\
 & & & & & &  0651330701  &  W1, M2, W2\\
 & & & & & &  0651331101  &  W1, M2, W2\\
PG1116+215  &  169.786 &  21.3216 &  0.1765 &  8.53 &  0.0186 &  0201940101  &  U, B, V, W1, M2, \\
 & & & & & &  0201940201  &  M2, \\
 & & & & & &  0554380201  &  U, W1, M2, W2\\
 & & & & & &  0554380301  &  U, W1, M2, W2\\
 & & & & & &  0554380101  &  U, W1, M2, W2\\
PG1351+640  &  208.316 &  63.7626 &  0.0882 &  7.66 &  0.0313 &  0205390301  &  U, B, V, W1, M2, \\
 & & & & & &  0556230201  &  W1, M2, W2\\
 & & & & & &  0556230101  &  W1, M2, \\
PG1402+261  &  211.318 &  25.9261 &  0.1640 &  7.94 &  0.0214 &  0400200201  &  U, W1, M2, W2\\
 & & & & & &  0400200101  &  U, W1, M2, W2\\
PG1440+356  &  220.531 &  35.4397 &  0.0790 &  7.47 &  0.0150 &  0005010301  &  W1, W2\\
 & & & & & &  0107660201  &  U, V, W1, M2, W2\\
 & & & & & &  0005010201  &  W1, W2\\
 & & & & & &  0005010101  &  W1, W2\\
Q0056-363  &  14.6556 &  -36.1013 &  0.1641 &  8.95 &  0.0281 &  0205680101  &  U, W1, M2, W2\\
 & & & & & &  0401930101  &  W1, M2, W2\\
 & & & & & &  0102040701  &  W2\\
RE1034+396  &  158.661 &  39.6411 &  0.0424 &  6.41 &  0.0214 &  0506440101  &  U, B, V, W1, W2\\
 & & & & & &  0675440201  &  W1, \\
 & & & & & &  0675440101  &  W1, \\
 & & & & & &  0109070101  &  U, W1, M2, W2\\
 & & & & & &  0655310101  &  W1, \\
 & & & & & &  0675440301  &  W1, \\
 & & & & & &  0561580201  &  W1, \\
 & & & & & &  0655310201  &  W1, \\
UGC3973  &  115.637 &  49.8096 &  0.0221 &  7.72 &  0.0829 &  0103862101  &  W2\\
 & & & & & &  0400070401  &  W1, W2\\
 & & & & & &  0502091001  &  B, W1, M2, W2\\
 & & & & & &  0400070201  &  W1, W2\\
 & & & & & &  0400070301  &  W1, W2\\

\caption{List of objects of our sample. The reddening E(B-V) is calculated from the Galactic extinction following \cite{guv09} unless specified\label{tab1}}\\
\end{longtable}
}
\newpage
\section{The best fit parameter and bolomeric luminosity values}
\label{appbestfit}
\begin{longtable}{cccccccc}
\hline\hline
Source  & ObsID & $\Gamma_{wc}$ & $kT_{wc}$ & $\Gamma_{hc}$ & $kT_{bb}$ & $\chi^2_{red}$ & dof \\
 & & & (keV) & & (eV) & & \\
\hline
\endfirsthead
\caption{continued.}\\
\hline\hline
Source  & ObsID & $\Gamma_{wc}$ & $kT_{wc}$ & $\Gamma_{hc}$ & $kT_{bb}$ & $\chi^2_{red}$ & dof \\
 & & & (keV) & & (eV) & &\\
\hline
\endhead
\hline
\endfoot
1H0419-577 & 0148000201 & 2.48$\pm$ 0.17 & 0.16$\pm$ 0.01 & 1.50$\pm$ 0.04 & 3.5$\pm$ 1.4 & 1.29 & 196\\
 & 0148000301 & 2.84$\pm$ 0.44 & 0.14$\pm$ 4.95 & 1.76$\pm$ 0.08 & 4.5$\pm$ 2.9 & 1.32 & 175\\
 & 0148000401 & 3.22$\pm$ 0.16 & 0.30$\pm$ 0.08 & 1.77$\pm$ 0.06 & 13.3$\pm$ 2.0 & 0.98 & 201\\
 & 0148000501 & 3.01$\pm$ 0.19 & 0.44$\pm$ 4.85 & 1.68$\pm$ 0.13 & 7.3$\pm$ 2.3 & 1.13 & 197\\
 & 0148000601 & 2.74$\pm$ 0.12 & 0.75$\pm$ 4.85 & 1.63$\pm$ 0.13 & 3.9$\pm$ 0.8 & 0.96 & 197\\
 & 0604720301 & 2.72$\pm$ 0.03 & 0.41$\pm$ 0.03 & 1.61$\pm$ 0.03 & 5.8$\pm$ 0.3 & 1.29 & 221\\
 & 0604720401 & 2.65$\pm$ 0.04 & 0.45$\pm$ 0.06 & 1.59$\pm$ 0.04 & 4.4$\pm$ 0.3 & 1.10 & 221\\
 & 0112600401 & 2.90$\pm$ 0.14 & 0.34$\pm$ 0.13 & 1.85$\pm$ 0.08 & 6.3$\pm$ 1.2 & 0.84 & 181\\
ESO198-G24 & 0305370101 & 2.53$\pm$ 0.02 & 0.31$\pm$ 0.02 & 1.72$\pm$ 0.02 & 3.0 (f) & 1.55 & 226\\
 & 0112910101 & 2.42$\pm$ 0.09 & 0.27$\pm$ 0.03 & 1.85$\pm$ 0.04 & - & 1.16 & 184\\
 & 0067190101 & 2.44$\pm$ 0.04 & 0.23$\pm$ 0.01 & 1.83$\pm$ 0.01 & - & 1.30 & 222\\
HB890405-123 & 0202210401 & 2.79$\pm$ 0.08 & 0.31$\pm$ 0.06 & 1.76$\pm$ 0.02 & 2.1$\pm$ 0.2 & 1.10 & 219\\
 & 0202210301 & 2.76$\pm$ 0.03 & 0.31$\pm$ 0.02 & 1.75$\pm$ 0.03 & 2.1$\pm$ 0.1 & 0.96 & 218\\
HE1029-1401 & 0203770101 & 2.66$\pm$ 0.01 & 0.38$\pm$ 0.03 & 1.81$\pm$ 0.03 & 2.5 (f) & 1.33 & 226\\
 & 0110950101 & 2.75$\pm$ 0.05 & 0.34$\pm$ 0.11 & 1.95$\pm$ 0.09 & - & 1.06 & 168\\
IRASF12397+3333 & 0202180301 & 2.31$\pm$ 0.19 & 0.36$\pm$ 0.10 & 2.06$\pm$ 0.14 & 8.0$\pm$ 3.4 & 0.99 & 159\\
 & 0202180201 & 2.33$\pm$ 0.06 & 0.19$\pm$ 0.01 & 2.30$\pm$ 0.03 & 20.0$\pm$ 0.5 & 1.32 & 226\\
LBQS1228+1116 & 0306630201 & 2.71$\pm$ 0.06 & 0.45$\pm$ 0.08 & 1.86$\pm$ 0.08 & 2.5$\pm$ 0.3 & 1.20 & 191\\
 & 0306630101 & 2.68$\pm$ 0.07 & 0.42$\pm$ 0.08 & 1.84$\pm$ 0.10 & 2.3$\pm$ 0.4 & 1.06 & 177\\
MRK279 & 0302480501 & 2.45$\pm$ 0.03 & 0.24$\pm$ 0.01 & 1.79$\pm$ 0.02 & 2.8$\pm$ 0.3 & 1.31 & 224\\
 & 0302480601 & 2.57$\pm$ 0.07 & 0.30$\pm$ 0.05 & 1.86$\pm$ 0.03 & 3.4$\pm$ 0.6 & 0.98 & 224\\
 & 0302480401 & 2.54$\pm$ 0.04 & 0.32$\pm$ 0.03 & 1.85$\pm$ 0.02 & 3.3$\pm$ 0.4 & 1.37 & 224\\
MRK335 & 0600540601 & 2.43$\pm$ 0.04 & 0.15$\pm$ 0.00 & 1.94$\pm$ 0.03 & 2.6$\pm$ 0.3 & 1.50 & 224\\
 & 0600540501 & 2.40$\pm$ 0.05 & 0.16$\pm$ 0.01 & 1.96$\pm$ 0.03 & 2.8$\pm$ 0.4 & 1.41 & 220\\
 & 0510010701 & 2.76$\pm$ 0.07 & 0.26$\pm$ 0.04 & 1.52$\pm$ 0.06 & 2.2$\pm$ 0.3 & 1.74 & 172\\
MRK509 & 0601390201 & 2.74$\pm$ 0.03 & 0.57$\pm$ 0.07 & 1.75$\pm$ 0.03 & 4.0$\pm$ 0.2 & 1.06 & 225\\
 & 0601390301 & 2.67$\pm$ 0.02 & 0.55$\pm$ 0.05 & 1.76$\pm$ 0.03 & 3.5$\pm$ 0.2 & 1.33 & 225\\
 & 0601390401 & 2.80$\pm$ 0.03 & 0.65$\pm$ 0.12 & 1.77$\pm$ 0.04 & 5.0$\pm$ 0.2 & 1.24 & 225\\
 & 0601390501 & 2.86$\pm$ 0.02 & 0.61$\pm$ 0.07 & 1.72$\pm$ 0.03 & 5.6$\pm$ 0.2 & 1.49 & 225\\
 & 0601390601 & 2.85$\pm$ 0.02 & 0.48$\pm$ 0.04 & 1.80$\pm$ 0.03 & 6.2$\pm$ 0.2 & 1.46 & 225\\
 & 0601390701 & 2.75$\pm$ 0.02 & 0.50$\pm$ 0.05 & 1.78$\pm$ 0.03 & 4.6$\pm$ 0.2 & 1.76 & 225\\
 & 0601390801 & 2.70$\pm$ 0.02 & 0.56$\pm$ 0.06 & 1.77$\pm$ 0.03 & 4.2$\pm$ 0.2 & 1.32 & 225\\
 & 0601390901 & 2.63$\pm$ 0.02 & 0.55$\pm$ 0.05 & 1.78$\pm$ 0.03 & 3.6$\pm$ 0.2 & 1.24 & 225\\
 & 0601391001 & 2.70$\pm$ 0.02 & 0.44$\pm$ 0.03 & 1.80$\pm$ 0.02 & 4.1$\pm$ 0.2 & 1.51 & 225\\
 & 0601391101 & 2.68$\pm$ 0.02 & 0.48$\pm$ 0.04 & 1.76$\pm$ 0.02 & 3.9$\pm$ 0.2 & 1.35 & 225\\
 & 0306090101 & 2.58$\pm$ 0.16 & 0.62$\pm$ 4.80 & 1.71$\pm$ 0.16 & 2.9$\pm$ 0.9 & 1.07 & 191\\
 & 0306090201 & 2.67$\pm$ 0.02 & 0.55$\pm$ 0.05 & 1.72$\pm$ 0.03 & 5.0$\pm$ 0.2 & 1.20 & 222\\
 & 0306090301 & 2.68$\pm$ 0.03 & 0.43$\pm$ 0.04 & 1.78$\pm$ 0.03 & 5.1$\pm$ 0.3 & 1.24 & 223\\
 & 0306090401 & 2.63$\pm$ 0.03 & 0.42$\pm$ 0.03 & 1.74$\pm$ 0.02 & 3.7$\pm$ 0.2 & 1.35 & 223\\
 & 0130720101 & 2.60$\pm$ 0.01 & 0.32$\pm$ 0.05 & 1.70$\pm$ 0.03 & $<$2.5 & 1.20 & 222\\
 & 0130720201 & 2.63$\pm$ 0.04 & 0.38$\pm$ 0.05 & 1.78$\pm$ 0.03 & 3.4$\pm$ 0.3 & 1.67 & 223\\
MRK590 & 0109130301 & 2.37$\pm$ 0.23 & 0.47$\pm$ 4.84 & 1.68$\pm$ 0.18 & 1.1$\pm$ 0.8 & 0.83 & 144\\
 & 0201020201 & 2.76$\pm$ 0.44 & 0.40$\pm$ 4.98 & 1.77$\pm$ 0.03 & 1.0$\pm$ 0.1 & 1.17 & 225\\
MRK883 & 0652550201 & 2.32$\pm$ 0.08 & 0.17$\pm$ 0.03 & 1.87$\pm$ 0.11 & 1.0$\pm$ 0.1 & 1.17 & 128\\
 & 0302260701 & 2.21$\pm$ 0.08 & 0.18$\pm$ 0.03 & 1.74$\pm$ 0.14 & 1.0$\pm$ 0.2 & 1.26 & 104\\
 & 0302261001 & 2.14$\pm$ 0.21 & 0.16$\pm$ 0.11 & 1.91$\pm$ 0.23 & 1.0$\pm$ 0.2 & 1.16 & 114\\
 & 0302260101 & 2.37$\pm$ 0.10 & 0.31$\pm$ 0.16 & 1.50$\pm$ 0.06 & 1.0$\pm$ 0.2 & 1.26 & 95\\
NGC4593 & 0740920201 & 2.27$\pm$ 0.04 & 0.76$\pm$ 0.15 & 1.50$\pm$ 0.10 & 1.7 (f) & 1.61 & 151\\
 & 0740920301 & 2.54$\pm$ 0.04 & 0.15$\pm$ 0.03 & 1.69$\pm$ 0.03 & - & 1.13 & 143\\
 & 0740920401 & 2.44$\pm$ 0.03 & 0.16$\pm$ 0.02 & 1.78$\pm$ 0.03 & - & 1.45 & 144\\
 & 0740920501 & 2.30$\pm$ 0.02 & 0.13$\pm$ 0.01 & 1.84$\pm$ 0.02 & - & 1.34 & 149\\
 & 0740920601 & 2.35$\pm$ 0.02 & 0.14$\pm$ 0.01 & 1.88$\pm$ 0.02 & - & 1.34 & 154\\
 & 0109970101 & 2.06$\pm$ 0.03 & 0.12$\pm$ 0.01 & 1.87$\pm$ 0.02 & - & 1.27 & 222\\
 & 0059830101 & 2.35$\pm$ 0.01 & 0.14$\pm$ 0.00 & 1.89$\pm$ 0.01 & - & 2.17 & 224\\
PG0804+761 & 0605110101 & 2.67$\pm$ 0.05 & 0.27$\pm$ 0.03 & 2.00$\pm$ 0.07 & 2.5$\pm$ 0.3 & 1.18 & 216\\
 & 0605110201 & 2.77$\pm$ 0.04 & 0.90$\pm$ 0.33 & 1.50$\pm$ 0.20 & 3.0$\pm$ 0.2 & 1.33 & 211\\
PG0844+349 & 0554710101 & 2.70$\pm$ 0.17 & 0.17$\pm$ 0.02 & 1.50$\pm$ 0.02 & 2.6$\pm$ 1.0 & 1.05 & 85\\
 & 0103660201 & 2.52$\pm$ 0.13 & 0.23$\pm$ 0.04 & 2.13$\pm$ 0.07 & 2.4$\pm$ 0.9 & 1.14 & 182\\
PG1114+445 & 0651330201 & 2.00$\pm$ 0.06 & 0.10$\pm$ 0.01 & 1.76$\pm$ 0.14 & 2.6 (f) & 0.82 & 115\\
 & 0651330101 & 2.24$\pm$ 0.13 & 0.10$\pm$ 0.01 & 1.51$\pm$ 0.05 & - & 1.09 & 158\\
 & 0651330301 & 2.21$\pm$ 0.25 & 0.12$\pm$ 0.02 & 1.68$\pm$ 0.13 & - & 1.16 & 140\\
 & 0651330401 & 2.00$\pm$ 0.03 & 0.10$\pm$ 0.01 & 1.96$\pm$ 0.09 & - & 1.10 & 168\\
 & 0651330601 & 2.13$\pm$ 0.02 & 0.11$\pm$ 0.00 & 1.76$\pm$ 0.06 & - & 1.11 & 177\\
 & 0651330901 & 2.25$\pm$ 0.10 & 0.11$\pm$ 0.01 & 1.74$\pm$ 0.09 & - & 1.09 & 171\\
 & 0651330801 & 2.30$\pm$ 0.14 & 0.11$\pm$ 0.01 & 1.65$\pm$ 0.07 & - & 1.22 & 170\\
 & 0651331001 & 2.39$\pm$ 0.11 & 0.12$\pm$ 0.01 & 1.67$\pm$ 0.08 & - & 0.82 & 164\\
 & 0651330501 & 2.20$\pm$ 0.18 & 0.13$\pm$ 0.01 & 1.84$\pm$ 0.10 & - & 1.11 & 153\\
 & 0109080801 & 2.24$\pm$ 0.08 & 0.11$\pm$ 0.01 & 1.87$\pm$ 0.05 & - & 0.88 & 191\\
 & 0651330701 & 2.24$\pm$ 0.07 & 0.12$\pm$ 0.01 & 1.70$\pm$ 0.08 & - & 0.98 & 171\\
 & 0651331101 & 2.21$\pm$ 0.02 & 0.11$\pm$ 0.01 & 1.60$\pm$ 0.05 & - & 1.09 & 151\\
PG1116+215 & 0201940101 & 2.85$\pm$ 0.04 & 0.35$\pm$ 0.03 & 1.92$\pm$ 0.05 & 3.3$\pm$ 0.4 & 1.34 & 225\\
 & 0201940201 & 2.74$\pm$ 0.07 & 0.30$\pm$ 0.06 & 2.09$\pm$ 0.13 & 1.0$\pm$ 0.5 & 1.06 & 109\\
 & 0554380201 & 2.87$\pm$ 0.02 & 0.30$\pm$ 0.01 & 1.84$\pm$ 0.03 & 3.6$\pm$ 0.1 & 1.27 & 218\\
 & 0554380301 & 2.99$\pm$ 0.03 & 0.35$\pm$ 0.01 & 1.85$\pm$ 0.04 & 4.3$\pm$ 0.4 & 1.24 & 209\\
 & 0554380101 & 2.86$\pm$ 0.03 & 0.44$\pm$ 0.06 & 1.84$\pm$ 0.03 & 4.0$\pm$ 0.3 & 1.10 & 218\\
PG1351+640 & 0205390301 & 2.78$\pm$ 0.09 & 0.24$\pm$ 0.04 & 2.01$\pm$ 0.12 & 2.1$\pm$ 0.4 & 1.11 & 130\\
 & 0556230201 & 2.58$\pm$ 0.26 & 0.15$\pm$ 0.03 & 1.50$\pm$ 0.10 & 2.2$\pm$ 1.1 & 1.08 & 46\\
 & 0556230101 & 2.24$\pm$ 0.19 & 0.11$\pm$ 0.02 & 1.50$\pm$ 0.08 & 1.9$\pm$ 1.0 & 1.79 & 25\\
PG1402+261 & 0400200201 & 2.90$\pm$ 0.22 & 0.64$\pm$ 4.84 & 1.86$\pm$ 0.32 & 3.2$\pm$ 1.2 & 1.08 & 112\\
 & 0400200101 & 2.83$\pm$ 0.13 & 0.31$\pm$ 0.10 & 2.04$\pm$ 0.13 & 3.4$\pm$ 0.9 & 0.97 & 164\\
PG1440+356 & 0005010301 & 2.59$\pm$ 0.10 & 0.18$\pm$ 0.03 & 2.33$\pm$ 0.09 & 2.1$\pm$ 0.8 & 1.09 & 141\\
 & 0107660201 & 2.51$\pm$ 0.10 & 0.19$\pm$ 0.02 & 2.39$\pm$ 0.08 & 2.5$\pm$ 0.6 & 0.78 & 169\\
 & 0005010201 & 2.68$\pm$ 0.18 & 0.21$\pm$ 0.07 & 2.22$\pm$ 0.15 & 3.6$\pm$ 1.2 & 1.09 & 163\\
 & 0005010101 & 2.65$\pm$ 0.19 & 0.20$\pm$ 0.05 & 2.29$\pm$ 0.09 & 3.5$\pm$ 1.4 & 1.30 & 174\\
Q0056-363 & 0205680101 & 2.71$\pm$ 0.06 & 1.83$\pm$ 4.58 & 1.51$\pm$ 0.12 & 3.2$\pm$ 0.4 & 1.29 & 216\\
 & 0401930101 & 2.62$\pm$ 0.09 & 0.34$\pm$ 0.08 & 1.95$\pm$ 0.07 & 2.0$\pm$ 0.6 & 0.98 & 191\\
 & 0102040701 & 2.68$\pm$ 0.10 & 0.32$\pm$ 0.09 & 2.08$\pm$ 0.11 & 1.0$\pm$ 0.7 & 0.85 & 146\\
RE1034+396 & 0506440101 & 2.82$\pm$ 0.05 & 0.19$\pm$ 0.01 & 2.39$\pm$ 0.04 & 12.9$\pm$ 0.5 & 1.90 & 169\\
 & 0675440201 & 3.20$\pm$ 0.11 & 0.23$\pm$ 0.03 & 2.01$\pm$ 0.20 & 20.0$\pm$ 0.5 & 0.83 & 117\\
 & 0675440101 & 3.16$\pm$ 0.05 & 0.21$\pm$ 0.02 & 2.12$\pm$ 0.18 & 20.0$\pm$ 0.3 & 1.16 & 117\\
 & 0109070101 & 3.02$\pm$ 0.10 & 0.21$\pm$ 0.03 & 2.28$\pm$ 0.13 & 17.9$\pm$ 1.4 & 1.38 & 90\\
 & 0655310101 & 3.14$\pm$ 0.08 & 0.20$\pm$ 0.02 & 2.26$\pm$ 0.14 & 19.7$\pm$ 0.7 & 1.01 & 121\\
 & 0675440301 & 2.50$\pm$ 0.08 & 0.16$\pm$ 0.01 & 2.11$\pm$ 0.12 & 14.3$\pm$ 1.4 & 1.15 & 118\\
 & 0561580201 & 2.61$\pm$ 0.07 & 0.15$\pm$ 0.01 & 2.35$\pm$ 0.11 & 15.7$\pm$ 1.1 & 1.14 & 130\\
 & 0655310201 & 3.19$\pm$ 0.10 & 0.22$\pm$ 0.04 & 2.17$\pm$ 0.18 & 19.3$\pm$ 0.9 & 1.13 & 141\\
UGC3973 & 0103862101 & 2.21$\pm$ 0.06 & 0.24$\pm$ 0.03 & 1.84$\pm$ 0.10 & 1.9 (f) & 1.16 & 160\\
 & 0400070401 & 2.30$\pm$ 0.02 & 0.17$\pm$ 0.01 & 1.90$\pm$ 0.03 & - & 1.35 & 219\\
 & 0502091001 & 2.39$\pm$ 0.03 & 0.19$\pm$ 0.01 & 1.50$\pm$ 0.00 & - & 2.36 & 225\\
 & 0400070201 & 2.17$\pm$ 0.04 & 0.16$\pm$ 0.01 & 2.09$\pm$ 0.03 & - & 1.32 & 223\\
 & 0400070301 & 2.42$\pm$ 0.04 & 0.37$\pm$ 0.07 & 1.82$\pm$ 0.05 & - & 1.09 & 219\\

\caption{Best fit parameters of each observation of our sample. The (f) means that the parameter is frozen during the fitting procedure. \label{bestfit}}\\
\end{longtable}

\begin{table}
\begin{tabular}{cc}
\begin{minipage}{0.5\textwidth}
\begin{tabular}{ccc}
\hline\hline
Source  & ObsID & $L_{bol}/L_{Edd}$ \\
\hline
1H0419-577 & 0148000201 &  0.098\\
 & 0148000301 &  0.097\\
 & 0148000401 &  0.290\\
 & 0148000501 &  0.166\\
 & 0148000601 &  0.102\\
 & 0604720301 &  0.156\\
 & 0604720401 &  0.128\\
 & 0112600401 &  0.195\\
 & Average &  0.154\\
ESO198-G24 & 0305370101 &  0.012\\
 & 0112910101 &  0.009\\
 & 0067190101 &  0.013\\
 & Average &  0.011\\
HE1029-1401 & 0203770101 &  0.102\\
 & 0110950101 &  0.090\\
 & Average &  0.096\\
IRASF12397+3333 & 0202180201 &  0.615\\
 & 0202180301 &  0.270\\
 & Average &  0.443\\
MRK279 & 0302480501 &  0.120\\
 & 0302480601 &  0.121\\
 & 0302480401 &  0.127\\
 & Average &  0.123\\
MRK335 & 0600540601 &  0.172\\
 & 0600540501 &  0.186\\
 & 0510010701 &  0.167\\
 & Average &  0.175\\
MRK509 & 0601390201 &  0.127\\
 & 0601390301 &  0.123\\
 & 0601390401 &  0.157\\
 & 0601390501 &  0.176\\
 & 0601390601 &  0.197\\
 & 0601390701 &  0.157\\
 & 0601390801 &  0.146\\
 & 0601390901 &  0.131\\
 & 0601391001 &  0.135\\
 & 0601391101 &  0.134\\
 & 0306090101 &  0.068\\
 & 0306090201 &  0.093\\
 & 0306090301 &  0.097\\
 & 0306090401 &  0.092\\
 & 0130720101 &  0.056\\
 & 0130720201 &  0.077\\
 & Average &  0.123\\
MRK590 & 0109130301 &  0.008\\
 & 0201020201 &  0.009\\
 & Average &  0.008\\
MRK883 & 0652550201 &  0.044\\
 & 0302260701 &  0.036\\
 & 0302261001 &  0.039\\
 & 0302260101 &  0.035\\
 & Average &  0.038\\
NGC4593 & 0740920201 &  0.046\\
 & 0740920301 &  0.025\\
 & 0740920401 &  0.029\\
 & 0740920501 &  0.044\\
 & 0740920601 &  0.043\\
 & 0109970101 &  0.053\\
 & 0059830101 &  0.075\\
 & Average &  0.045\\
\end{tabular}
\end{minipage} &

\begin{minipage}{0.5\textwidth}
\vspace*{-1.4cm}
\begin{tabular}{ccc}
\hline\hline
Source  & ObsID & $L_{bol}/L_{Edd}$ \\
\hline
PG0804+761 & 0605110101 &  0.402\\
 & 0605110201 &  0.434\\
 & Average &  0.418\\
PG0844+349 & 0554710101 &  0.085\\
 & 0103660201 &  0.107\\
 & Average &  0.096\\
PG1114+445 & 0651330201 &  0.072\\
 & 0651330101 &  0.053\\
 & 0651330301 &  0.056\\
 & 0651330401 &  0.072\\
 & 0651330601 &  0.064\\
 & 0651330901 &  0.050\\
 & 0651330801 &  0.047\\
 & 0651331001 &  0.044\\
 & 0651330501 &  0.061\\
 & 0109080801 &  0.046\\
 & 0651330701 &  0.053\\
 & 0651331101 &  0.056\\
 & Average &  0.056\\
PG1116+215 & 0201940101 &  0.384\\
 & 0201940201 &  0.392\\
 & 0554380201 &  0.373\\
 & 0554380301 &  0.425\\
 & 0554380101 &  0.404\\
 & Average &  0.395\\
PG1351+640 & 0205390301 &  0.269\\
 & 0556230201 &  0.238\\
 & 0556230101 &  0.296\\
 & Average &  0.268\\
PG1402+261 & 0400200201 &  0.438\\
 & 0400200101 &  0.485\\
 & Average &  0.461\\
PG1440+356 & 0005010301 &  0.458\\
 & 0107660201 &  0.635\\
 & 0005010201 &  0.625\\
 & 0005010101 &  0.635\\
 & Average &  0.588\\
Q0056-363 & 0205680101 &  0.053\\
 & 0401930101 &  0.041\\
 & 0102040701 &  0.073\\
 & Average &  0.056\\
RE1034+396 & 0506440101 &  1.500\\
 & 0675440201 &  1.819\\
 & 0675440101 &  1.889\\
 & 0109070101 &  1.519\\
 & 0655310101 &  1.804\\
 & 0675440301 &  1.691\\
 & 0561580201 &  1.698\\
 & 0655310201 &  1.750\\
 & Average &  1.709\\
UGC3973 & 0103862101 &  0.027\\
 & 0400070401 &  0.028\\
 & 0502091001 &  0.025\\
 & 0400070201 &  0.034\\
 & 0400070301 &  0.025\\
 & Average &  0.028\\
\end{tabular}
\end{minipage}
\end{tabular}
\caption{Bolometric luminosities (in Eddington unit) for the different observations of the sample for which an estimation of the object black hole mass is known (see Tab. \ref{tab1}). We have also reported the average bolometric luminosities for each object.  \label{bestfitlbol}}

\end{table}


\section{Particular model fitting procedure}
\label{secapp}
For 13 objects of our sample, HB890405-123, LBQS1228+1116, MRK 279, MRK 883, PG0804+761, PG0844+349, PG1116+215, PG1351+640, PG1402+261, PG1440+356, Q0056-363, RE1034+396 and PG1114+445 the fitting procedure explained in Sect. \ref{fitproc} gives good fits.  For the other objects however, additional model components significantly improve the fit. For IRASF12397+3333, MRK509, MRK590, ESO198-G24, HE1029-1401, NGC4593 and UGC3973 we add two narrow Gaussian lines to fit the strong residuals of emission/absorption line-like features below 2 keV. For 1H0419-577 and MRK335 we add a second CLOUDY table to fit the residual WA features.
In the case of PG1114+445, ESO198-G24, HE1029-1401, NGC4593 and UGC3973 the disk temperature is also fixed during the fitting procedure.

\label{secapp}
\end{appendix}

\end{document}